# Lunar Navigation System Optimization for Targeted Coverage with Semi-Analytical Station Keeping Model and Earth-GPS Integration

Rameez A. Malik[a,*], Yang Yang[a]

[a] School of Mechanical and Manufacturing Engineering, University of New South Wales, NSW, Sydney 2052, Australia

ARTICLE INFO

ABSTRACT

The design of an indigenous Lunar Navigation Satellite System (LNSS) is receiving growing attention due to the surge in planned lunar missions and the limited accessibility of Earth-based Global Navigation Satellite Systems (GNSS) in the cislunar environment. Several studies have explored LNSS architecture using geometric analysis in both near and distant lunar orbits. The existing LNSS optimization efforts have primarily focused on global lunar coverage using analytical station-keeping models with low accuracy. Furthermore, current south pole-focused research is restricted to Elliptical Lunar Frozen Orbits (ELFOs) and lacks comprehensive optimization approach. Additionally, integration with Earth GNSS systems for ephemeris computation and time synchronization has not been adequately addressed in prior studies. In this work, we present a comprehensive LNSS mission design framework based on evolutionary multi-objective optimization integrated with a high-fidelity numerical lunar orbit propagation model. The optimization simultaneously considers navigation performance in the lunar south pole region, semi-analytical continuous station-keeping maneuver model for realistic $\Delta V$ estimate, and GPS-LNSS integration analysis parameters. The resulting Pareto front offers a diverse set of LNSS configurations that balance coverage, accuracy, and $\Delta V$ requirements. Notably, the optimization identifies diverse non-frozen elliptical orbit solutions that achieve over 90% south pole coverage with acceptable navigation accuracy using as few as six satellites and $\Delta V$ of less than 0.4 $\mathrm{Kms^{-1}}$ per satellite per year. This represents a significant reduction in constellation size compared to previous studies, offering a cost-effective yet operationally efficient solution for future LNSS missions.

## List of Abbreviations

| | |
|---|---|
| AOP | Argument of Perilune |
| DOP | Dilution of Precision |
| DRO | Distant Retrograde Orbit |
| ECI | Earth Centered Inertial frame |
| EIRP | Effective Isotropic Radiated Power |
| ELFO | Elliptical lunar Frozen Orbit |
| ET | Ephemeris Time |
| FOMs | Figure of Merits |
| FSL | Free Space Loss |
| GDOP | Geometric Dilution of Precision |
| GEO | Geostationary Orbit |
| GNSS | Global Navigation Satellite System |
| GPS | Global Positioning System |
| GVEs | Gauss's Variational Equations |
| ICRF | International Celestial Reference Frame |
| LFOs | Lunar Frozen Orbits |
| LLOs | Low Lunar Orbits |
| LNSS | Lunar Navigation Satellite System |
| LOS | Line-of-Sight |
| LPOs | Lagrange Point Orbits |
| LVLH | Local Vertical Local Horizontal |
| MCI | Moon Centered Inertial frame |
| MCMF | Moon-Centered-Moon-Fixed frame |
| MOEA | Multi-Objective Evolutionary Algorithm |
| MOGA | Multi-Objective Genetic Algorithm |
| NASA | National Aeronautics and Space Administration |
| NLP | Non-Linear Programming |
| NSGA-II | Non-dominated Sorting Genetic Algorithm II |
| PDOP | Position Dilution of Precision |
| PNT | Positioning, Navigation, and Timing |
| RAAN | Right Ascension of the Ascending Node |
| RTN | Radial–Tangential–Normal |
| SMA | Semi-Major Axis |
| SPICE | Spacecraft Planet Instrument C-matrix Events |
| SQP | Sequential Quadratic Programming |
| SRP | Solar Radiation Pressure |
| TDB | Barycentric Dynamical Time |
| TLE | Two Line Element |
| UERE | User Equivalent Ranging Error |

# 1 Introduction

Lunar exploration has always been a crucial aspect of our space endeavors, serving as a steppingstone for deep space missions and enhancing our understanding of the solar system. However, latest authentic evidence of water ice availability in lunar poles has completely changed our insights about this Earth's natural satellite [1]. The international space agencies and commercial space organizations aim to have a permanent footprint on the Moon. The US National Aeronautics and Space Administration (NASA) has revealed its plan to have a prominent human presence on the Moon through Artemis missions in near future [1]. These advancements clearly indicate that there is huge potential in the lunar economy in the coming years. The global space organizations are making collaborative efforts to implement sustainable space tourism to support lunar missions with a high number of crew and lunar rovers. The future lunar missions will require reliable navigation and positioning infrastructure to efficiently carry out safe in-orbit, landing and surface operations. Current research on exploring the capacity of Earth-based Global Navigation Satellite System (GNSS) signal reception and utilization in the cislunar space have shown promising prospects [3]. However, they also identified view geometry constraints that limit coverage and positional accuracy especially in Low Lunar Orbits (LLOs) and surface. For instance, the Earth-based GNSS provides about 98.8% service availability for lunar transfer trajectories with a massive drop to 11.6% on reaching the lunar orbits [4]. The feasible solution to this problem is to design a dedicated Lunar Navigation Satellite System (LNSS) for reliability of mission during all critical operational phases, e.g. descent and landing and surface operations, etc.

The concept of establishing a dedicated lunar navigation system has garnered significant attention in recent years, prompting numerous studies and research efforts in this domain. Wang et al. [5] proposed an analytical geometric framework for designing an LNSS in distant lunar orbits. Their study relies on visibility analysis and Position Dilution of Precision (PDOP) as key performance metrics to achieve global lunar coverage. The constellation architecture consists of 16 satellites placed in a combination of Halo and Distant Retrograde Orbit (DRO), offering a novel for sustained lunar navigation support. On the other hand, Gao et al. [6] explored a hybrid navigation and communication system based on Lagrange Point Orbits (LPOs) using an analytical geometric approach. This study is primarily focused on satellite visibility and surface coverage across various orbital configurations. Their proposed system comprising five satellites (three in DRO and two in LPO) is sufficient to achieve 100% coverage of the lunar surface. These prior studies only assess the performance on coverage aspects, ignoring other critical mission design and operation factors, e.g. communication constraints, integration with Earth-based navigation systems for time synchronization, etc. Conti et al [7] come up with a more comprehensive lunar navigation and communication mission design in stable and unstable halo orbital domain. The system is evaluated for lunar navigation segment for global lunar users, Earth-Moon communication and station keeping $\Delta V$ budget. A system of comprising L1, L2 and L3 orbits is proposed to meet the navigation and communication requirements.

Lunar Frozen Orbits (LFOs) have attracted considerable attention due to their relative stability and reduced sensitivity to lunar mascon perturbations. Wang et al have in [8] come with a more comprehensive mission design with a muti-objective approach in which Earth-based Beidou system is integrated with a hybrid LNSS in Elliptical lunar Frozen Orbit (ELFO) and Halo orbital configuration. The system is evaluated for autonomous orbit determination and time synchronization along navigation accuracy of lunar transfer trajectories and south pole user. Similarly, hybrid lunar navigation and communication constellation is proposed by Bhamidipati et al [9] based on small satellite and low-grade chip scale atomic clock. They have also explored three ELFO configurations and tested the system for User Equivalent Ranging Error (UERE), DOP and timing accuracy along with trade-off analysis on system size, weight and power requirements. Conti et al. [10] investigated LFOs for potential lunar navigation and communication applications using an averaged dynamical model that accounts for the perturbing influence of the Earth, assumed to lie on the lunar equatorial plane and the effect of lunar oblateness. Their results indicate that these orbits are not perfectly frozen and exhibit limited variations in orbital elements due to additional perturbations present in the cislunar environment. However, by appropriately selecting the combination of inclination, eccentricity, and Argument of Perilune (AOP), it is possible to achieve quasi-stable orbital configurations that meet the stability and coverage requirements for future lunar Positioning, Navigation, and Timing (PNT) missions. It is important to note that the frozen orbit assumption inherently constrains certain orbital elements thereby limiting the search space available for constellation optimization. Moreover, despite their quasi-stable nature, LFOs still experience gradual orbital drifts under realistic perturbations from the third body effect and Solar Radiation Pressure (SRP), necessitating periodic station-keeping. Therefore, in the context of LNSS design, exploring both frozen and non-frozen orbital configurations can be equally valuable, provided that the resulting station-keeping budget remains within acceptable operational range.

*Corresponding author

rameez.malik@unsw.edu.au (Rameez. A Malik)

https://orcid.org/0000-0002-1053-27configuration

Geometric analysis alone is not a very viable method to select an optimum constellation, especially if the problem involves multi-dimensional design settings. It requires a sophisticated optimization approach, capable of handling highly dynamic design parameters (orbital configurations and objectives). There are several prior studies that specifically dealt with numerical optimization techniques in mission design and analysis of lunar navigation system.

For instance, Pereira et al. [11] investigated LNSS architecture design of LFOs by applying the Borg Multi-Objective Evolutionary Algorithm (MOEA). The fitness function constitutes Geometric Dilution of Precision (GDOP), space segment cost, and the operational $\Delta V$ budget of the constellation. The study provides insightful preliminary results but focuses exclusively on LFO-based configurations, thereby constraining the solution space within a narrow band of orbital parameters. Moreover, the $\Delta V$ maneuvers are computed analytically in the Local Vertical Local Horizontal (LVLH) frame using mean Keplerian orbital element theory. The narrow orbital geometric range constraint of this study is efficiently addressed by Arcia et al. [12]. In this study, optimization is performed with comparatively a broad range of orbital parameters, considering PDOP, HDOP and constellation size HDOP as primary Figure of Merits (FOMs) along with station-keeping $\Delta V$ as a supporting factor. This algorithm achieved considerable improvement by 44 % in PDOP measurements with same number of satellites as compared to previous study. A simple GA optimization is applied by Hartigan et al [13] on ELFO and NRHO based LNSS constellations. It is not an exclusive optimized system in which only true anomalies are optimized to enhance the coverage, PDOP and navigation error estimates.

The integration of LNSS with Earth-based GNSS is essential to satisfy several key operational requirements during both deployment and routine operations. Such integration supports precise LNSS ephemeris generation, accurate orbit determination, and reliable time synchronization with terrestrial reference frames. Since the Moon lacks pre-established navigation and timing infrastructure, Earth-based GNSS signals, when accessible in lunar orbit, can provide precise orbital estimation and timing corrections. This interoperability enables cross-domain navigation solutions, enhances the accuracy and robustness of lunar orbit determination, and ensures consistency between lunar and terrestrial time scales, which is fundamental for future lunar exploration, communication, and PNT services [14][15]. Bhamidipati et al [14] proposes a time-transfer technique from the Earth's Global Positioning System (GPS) system to the LNSS system and conducts an in depth analysis of their approach in simulated environment. The timing corrections are estimated with Kalman filter, and the results verify the usability of available GPS signals for time synchronization. the existing research on the GPS-LNSS integrity (time synchronization and LNSS ephemeris computation) is confined to standalone ELFO configuration. It is determined by the GPS-LNSS time dependent visibility in a dynamic setting which makes it a highly geometry sensitive parameter. For instance, there is a possibility that non-frozen LNSS configurations have a superior GPS-LNSS geometry providing better system visibility or PDOP.

The existing research on LNSS optimization has largely focused on enhancing navigation performance at a global scale, which typically necessitates a higher number of satellites. An LNSS with a large constellation size is not a favorable solution considering launch, deployment and operational costs. Lunar south pole has emerged as a key area of interest for the space community due to several compelling reasons, such as the presence of water ice, a rich concentration of scientifically valuable minerals, and the existence of the south pole Aitken basin, the largest and deepest impact crater in the solar system [16]. This region is considered vital for advancing long-term objectives in solar system exploration, including technological demonstrations and sustained human presence. Designing a targeted LNSS specifically optimized for the lunar south pole could substantially reduce the number of required satellites, thereby diminishing mission cost. Meanwhile, the station-keeping budget calculation of prementioned studies is based on a low accuracy analytical approach and therefore does not give conclusive constellation stability and maintenance analysis. For instance, the $\Delta V$ model in [11] incorporates maneuvers to correct the Semi-Major Axis (SMA), inclination, Right Ascension of the Ascending Node (RAAN), AOP, and eccentricity. While this model is computationally fast and addresses all key orbital corrections, it relies on mean Keplerian orbit parameters to compute the maneuvers. A limitation of this approach is its sensitivity to errors in these orbital parameters, i.e. if the actual spacecraft orbit deviates from the assumed mean values, the computed $\Delta V$ may be misapplied. It can lead to less effective corrections which may reduce robustness of the algorithm. Similarly, station-keeping $\Delta V$ in [12] incorporates only three orbital elements, i.e. SMA, eccentricity and inclination and is computed using Hohmann transfer and out-of-plane impulse burns. This is still a low precision analytical approach mainly adopted to reduce computational burden. Moreover, the LNSS-GPS integration has not been realized in the optimization frameworks in literature as one of the objective parameters in deciding LNSS geometry. In response to this background, this paper attempts to construct a more comprehensive LNSS optimization framework by considering compact mission design parameters covering the prementioned mission design aspects.

The optimization is handled with a Multi-Objective Genetic Algorithm (MOGA) with NSGA-II (Non-dominated Sorting Genetic Algorithm II) to deal with multiple conflicting objectives simultaneously. NSGA-II is an evolutionary algorithm that operates by maintaining a population of candidate solutions that improve over iterations / generations [17]. The solutions in each generation are ranked through fast non-dominated sorting into successive Pareto fronts based on dominance relations, while a crowding-distance metric preserves solution diversity across the front. The non-dominant solutions evolve across the generation depending on optimization control parameters. This approach is particularly advantageous for this research where competing performance metrics like positioning accuracy e.g., PDOP, HDOP, coverage availability, station-keeping $\Delta V$ and GPS-LNSS integrity parameters are optimized. The NSGA-II can explore high-dimensional, nonlinear, and non-convex solution spaces without requiring gradient information which makes it a robust choice for the complex dynamics [18]. The main contributions of this study are explained as follows:

1. This paper attempts to optimize LNSS with targeted coverage on lunar south pole to reduce mission design cost. Moreover, the minimization of the constellation size is treated as one of the optimization objectives to identify cost-effective solutions. However, the system with small number of spacecraft must be able to provide continuous positioning on the south pole. The positioning accuracy of a positioning system is mainly derived from the dilution of precision (DOP) [19]. There is no well-defined specification set to quantify the LNSS performance; however, the international space organizations have outlined key design criteria for a future LNCSS. For instance, the global exploration community has aimed 3D positioning accuracy for lunar surface users below 50 m and horizontal position up to 10 m 3-sigma measure [9].
2. This research diversifies the orbital search space in near lunar orbits to efficiently cover frozen and non-frozen orbits. The idea is to allow optimizer to explore diverse LNSS configurations that fulfil PNT needs with small number of satellites if orbital stability is acceptable.
3. The paper places particular emphasis on the station-keeping $\Delta V$ budget, which is evaluated using a continuous maneuver model based on the Gauss's Variational Equations (GVEs). Instead of using mean orbital elements over a specific interval, this approach computes actual instantaneous drift of all the secular and osculating orbital elements. The maneuver model then drives the instantaneous perturbed state of the LNSS satellites from the high-fidelity numerical lunar propagation model. The idea is to capture the dynamic perturbation effects which are embedded in the propagated state and are responsible for the drifting orbital elements. This approach enables reliable $\Delta V$ estimations as compared to those obtained through simplified analytical methods. Additionally, a maneuver optimization strategy is integrated within the main optimization framework to demonstrate idea to optimize thrust vector for effective resource utilization on-board in real world operations.
4. Incorporate GPS-LNSS integration as one of the objective variables to study the influence of the variation of LNSS geometry on the GPS-LNSS PDOP metric. The PDOP metric is mainly selected because it is primary measure for LNSS orbit determination. The whole idea is that if an LNSS configuration offers desired navigation performance and $\Delta V$ requirements are also satisfied, it should also be able to achieve optimum GPS-LNSS visibility and PDOP.

The organization of the paper is as follows: the mathematical model for computation of dilutions and coverage measures are explained in Section 2. The dynamic model including the high-fidelity lunar propagation parameters and the $\Delta V$ calculation logic has been thoroughly explained in Section 3. The simulation setup comprising the integration of optimization framework parameters with the performance metrics (objective function) and orbital geometry (decision variables) has been described in Section 5. Similarly, the simulation analyses of a standalone LNSS configuration to demonstrate computation process of the objective functions and MOGA-NSGS-II algorithm for complete LNSS mission design have been presented in Section 6. In addition, the influence of variation of orbital configuration on the objective functions has also been explained for the optimization trade-off. The conclusion and prospects of this study have been presented in Section 7.

# 2 Performance Metrics

## 2.1 Dilution of Precision

The positioning performance of the lunar navigation system is evaluated using DOP metrics, which quantify the effect of satellite-receiver geometry on the accuracy of position estimation. DOP is a dimensionless factor that scales measurement error, such that a lower DOP value indicates a more favorable satellite geometry and thus higher positioning accuracy. Among the various DOP metrics, PDOP and HDOP are most relevant metrics for assessing 3D and 2D location accuracy, respectively [20]. In this analysis, we incorporate only PDOP and HDOP to reduce computational complexity while maintaining sensitivity to geometric performance.

To evaluate the DOP metrics over the lunar south pole, we discretize the surface into a uniform latitude-longitude grid bounded between latitudes $\phi \in [-90°, -60°]$ and longitudes $\lambda \in [0°, 360°]$ with a resolution of 10° in lat/long direction. Let $\phi_i$ and $\lambda_i$ denote the latitude and longitude of the $i^{\text{th}}$ and $j^{\text{th}}$ grid coordinates, respectively. These are defined by:

$$\phi_i = \phi_{\min} + (i-1)\,\Delta\phi \tag{1}$$

$$\lambda_i = \lambda_{\min} + (j-1)\,\Delta\lambda \tag{2}$$

for $i = 1, \dots, n_\phi$, $j = 1, \dots, n_\lambda$, with $\phi_{\min} = -90°$, $\Delta\phi = 10$, $\lambda_{\min} = 0°$, $\Delta\lambda = 10°$. The total number of surface sampling points *is* $M = n_\phi \times n_\lambda$. At each epoch $t$ and grid node $(i, j)$, the visibility of satellite $k$ is determined by comparing its elevation angle $\theta_k^{(t,i,j)}$ against a minimum elevation mask of $\theta_{\min} = 5°$. We introduce the binary visibility indicator as follows:

$$\delta_k^{(t,i,j)} = \begin{cases} 1, \text{if } \theta_k^{(t,i,j)} \geq \theta_{\min} \\ 0, \quad \text{Otherwise} \end{cases} \tag{3}$$

It implies that the total number of visible satellites is:

$$N_v^{(t,i,j)} = \sum_{k=1}^{N_{\text{sat}}} \delta_k^{(t,i,j)} \tag{4}$$

The above equation succinctly captures per-grid point per-epoch satellite visibility count, which is then used to construct the geometry matrix only when $N_v^{(t,i,j)} \geq 4$. The satellite line-of-sight (LOS) unit vectors are used to construct the geometry matrix $G$. Each row of $G$ corresponds to the direction vector from the receiver to a visible satellite, expressed in the local reference frame:

$$G^{(t,i,j)} = \begin{bmatrix} \hat{l}_{x1}^{(t,i,j)} & \hat{l}_{y1}^{(t,i,j)} & \hat{l}_{z1}^{(t,i,j)} & 1 \\ \hat{l}_{x2}^{(t,i,j)} & \hat{l}_{y2}^{(t,i,j)} & \hat{l}_{z2}^{(t,i,j)} & 1 \\ \vdots & \vdots & \vdots & \vdots \\ \hat{l}_{xN_v}^{(t,i,j)} & \hat{l}_{yN_v}^{(t,i,j)} & \hat{l}_{zN_v}^{(t,i,j)} & 1 \end{bmatrix}, \tag{5}$$

here, $\hat{l}_{x1}^{(t,i,j)}$, $\hat{l}_{y1}^{(t,i,j)}$ and $\hat{l}_{z1}^{(t,i,j)}$ denote the components of the LOS unit vector from the receiver to the $i^{th}$ satellite, and the final column accounts for the receiver clock bias term, which is included in the formulation of $G$ but not explicitly analyzed in this work. The $G$ is used to form the information matrix $H$ as:

$$H = G^T G, \qquad H^{-1} = (G^T G)^\dagger \tag{6}$$

The inverse of $H$, denoted $H^{-1}$, encapsulates the positional uncertainty resulting purely from satellite geometry. From this, PDOP and HDOP are extracted using the diagonal elements:

$$PDOP_{(t,i,j)} = \sqrt{H_{11}^{-1} + H_{22}^{-1} + H_{33}^{-1}} \tag{7}$$

$$HDOP_{(t,i,j)} = \sqrt{H_{11}^{-1} + H_{22}^{-1}} \tag{8}$$

where, $H_{ij}^{-1}$ represents the elements in the i$^{\text{th}}$ row and j$^{\text{th}}$ column of $H^{-1}$. The statistical analysis of the dilutions is based on a 3-sigma filtering method that is applied to remove outliers from the time series data. It is important to note that clock bias is not modelled or estimated in this research, as the focus is strictly on geometric dilution metrics rather than full positioning solutions. Therefore, the DOP computations here reflect only the spatial geometry-induced dilution and are not affected by clock stability or timing error models. If *P* is the set of all observed PDOP values, we may compute the mean 3-sigma filtered DOPs:

$$P = \{PDOP_{(t,i,j)}\}, \qquad N = |P| = t_{\text{eph}} n_\phi n_\lambda, \tag{9}$$

$$\mu_P = \frac{1}{N} \sum_{t=1}^{t_{\text{eph}}} \sum_{i=1}^{n_\phi} \sum_{j=1}^{n_\lambda} PDOP_{(t,i,j)} \tag{10}$$

$$PDOP_{\sigma} = \sqrt{\frac{1}{N}\sum_{t=1}^{t_{\text{eph}}}\sum_{i=1}^{n_{\phi}}\sum_{j=1}^{n_{\lambda}}\left(PDOP_{(t,i,j)} - \mu_P\right)^2} \tag{11}$$

$$PDOP_{3\sigma} = \{p \in P \; |p - \mu_P| \leq 3PDOP_{\sigma}\} \tag{12}$$

$$\overline{PDOP_{3\sigma}} = \frac{1}{|PDOP_{3\sigma}|}\sum_{p \,\in\, PDOP_{3\sigma}} p \tag{13}$$

where $t_{\text{eph}}n_{\phi}n_{\lambda}$ is the number of spatiotemporal samples, $\mu_P$ is the mean of spatiotemporal $PDOP\ (P)$, $PDOP_{\sigma}$ is the standard deviation of P, $PDOP_{3\sigma}$ is the 3-sigma filtered PDOP set and $\overline{PDOP_{3\sigma}}$ is the mean of 3-sigma filtered PDOP. A similar procedure is applied to compute 3-sigma filtered HDOP ($\overline{HDOP_{3\sigma}}$). We have incorporated another navigation performance metric to quantify the availability of the LNSS in a compact manner. It is defined as the percentage of spatiotemporal grid points, covering all combinations of surface locations and time steps, where both PDOP and HDOP remain below 15. This threshold is selected because it falls within the acceptable category of DOP rankings. The DOP above 20 are considered poor and therefore should be excluded from DOP availability measure calculation. This measure ensures the availability metrics are computed using only high-quality and reliable DOP data. By systematically evaluating each grid point at every epoch, the simulation determines how frequently the navigation geometry supports reliable position estimation. The resulting availability metric thus provides a statistical overview of the system's capability to offer usable location accuracy throughout the evaluated region and over the full simulation duration. This is particularly critical for assessing LNSS performance in dynamic lunar environments, where satellite visibility and geometry vary significantly due to orbital motion and lunar rotation. It is mathematically expressed as:

$$PDOP_{\text{avail}} = \frac{\left|\{(t,i,j)\ PDOP_{t,i,j} < 15\}\right|}{t_{\text{eph}}n_{\phi}n_{\lambda}} \times 100\ \%,$$

$$HDOP_{\text{avail}} = \frac{\left|\{(t,i,j)\ HDOP_{t,i,j} < 15\}\right|}{t_{\text{eph}}n_{\phi}n_{\lambda}} \times 100\% \tag{14}$$

## 2.2 Earth GPS-LNSS Integration Model

As discussed before, The integration of LNSS with Earth-based navigation systems is crucial to meet several operational requirements, including LNSS ephemeris generation, orbit determination, and time-synchronization. Therefore, modeling of GPS-LNSS system Integration needs to be realistic for accurate analysis. The integration of both systems is governed by a combination of geometric and radio-frequency link constraints. In the presented model, a comprehensive visibility function is implemented that evaluates each GPS–LNSS LOS vector against three primary criteria at every epoch, i.e. lunar occultation, Earth occultation and signal strength (link budget). The GPS constellation is propagated with a two-body propagation model using the real-time Two Line Element (TLE) File data [21]. The computation of GPS visibility is performed in the MCI–J2000 reference frame, considering LOS vector from the LNSS to a GPS satellite which is defined as [22]:

$$r_{\text{LOS}} = r_{\text{GPS}} - r_{\text{LNSS}} \tag{15}$$

where $r_{\text{GPS}}$ and $r_{\text{LNSS}}$ are position vectors of the GPS and LNSS satellites, respectively. This LOS vector is normalized to obtain a unit vector in the direction of signal propagation:

$$\hat{u} = \frac{r_{\text{LOS}}}{\|r_{\text{LOS}}\|} \tag{16}$$

### 2.2.1 Lunar Occultation Model

In the first step, we need to compute lunar blockage for the LNSS satellite using the angular separation between the LOS direction and the local zenith i.e., the vector from the Moon's center to the LNSS. It is mathematically expressed using the dot product as [22][23]:

$$\theta_{\text{moon}} = \cos^{-1}(-\hat{u}.\hat{z}), \tag{17}$$

where $\hat{z} = \frac{r_{\text{LNSS}}}{\|r_{\text{LNSS}}\|}$ is the radius vector of the LNSS satellite. The Moon's angular radius, as seen from the LNSS, is expressed as:

$$\alpha_{\text{moon}} = \sin^{-1}\left(\frac{R_{\text{moon}}}{\|r_{\text{LNSS}}\|}\right) \tag{18}$$

where $R_{\text{moon}}$ is the physical radius of the Moon. If the LOS angle $\theta_{\text{moon}}$ falls below $\alpha_{\text{moon}}$, the GPS to LNSS signal path intersects the Moon's body, and hence the LOS is considered blocked. This constraint rigorously accounts for geometric lunar occultation, which is particularly prominent for the satellites near the lunar horizon as seen by the GPS satellite.

### 2.2.2 Earth Occultation Model

The Earth blockage modelling of the GPS signal can be described with help of GPS satellite view geometry configuration. For analysis, we have considered GPS Block IIR L1 C/A (Coarse/Acquisition) transmitter antenna with 70° off-boresight angle mask ($\theta_{\text{cone}}$), centered on Earth, as shown in Figure 1. The Figure illustrates the geometry associated with GPS signal availability in Geostationary Orbit (GEO). As per the geometry, GPS signals corresponding to off-boresight greater than 13.8° are able gaze by the limb of Earth and can be received on the opposite side of the Earth [24]. The feasibility of receiving GNSS signals in Earth-Moon L2 Halo orbit has been studied by Delépaut, A et al in reference [25]. This project explored the reception capacity of GPS and Galileo with a single standalone GNSS receiver having an acquisition threshold of 15 dB-Hz Carrier-to-Noise Power Spectral Density Ratio ($^{C}/_{N_0}$). The boresight vector, which is the orientation of the GPS antenna lobe, is taken as:

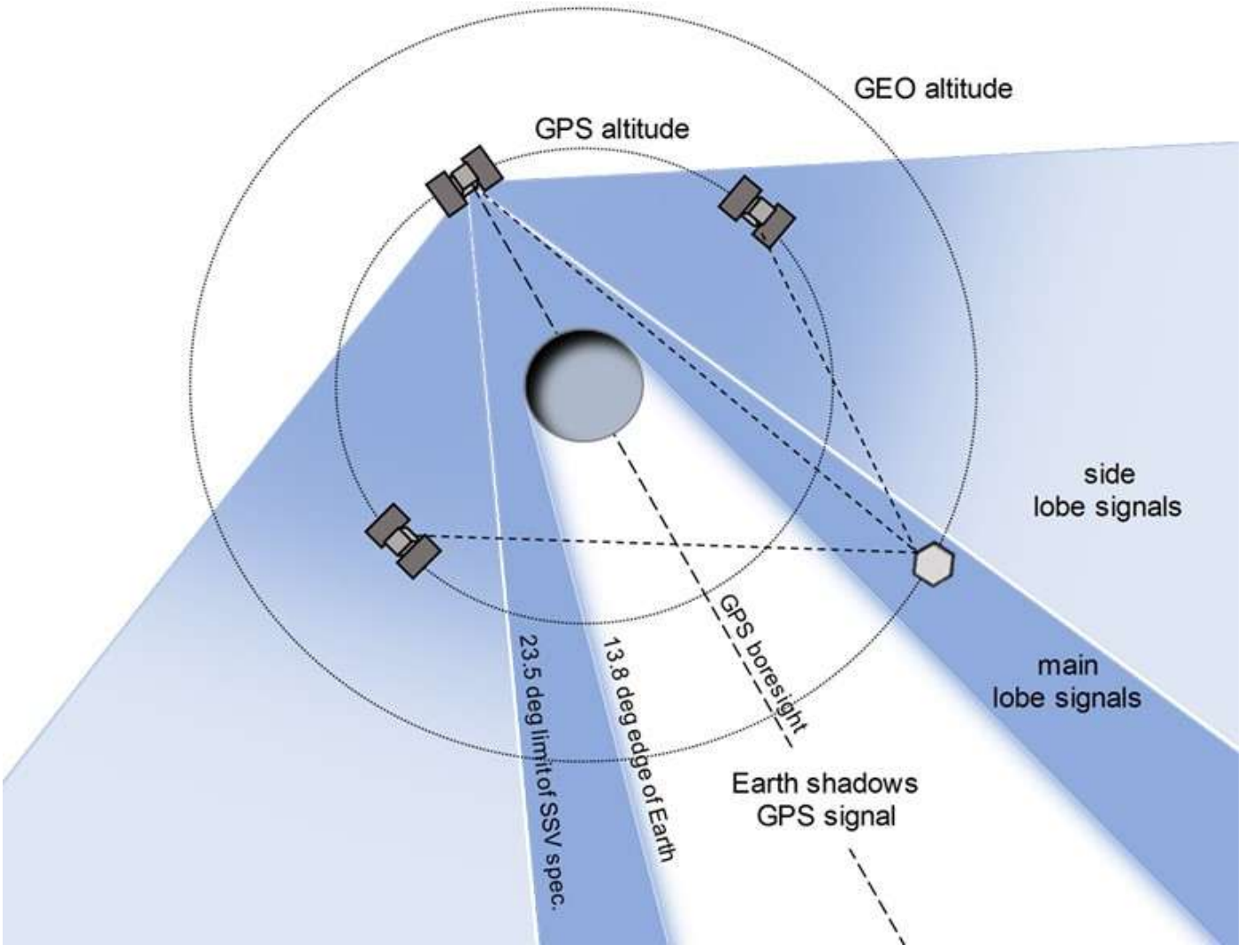

**Figure 1:** Illustration of GPS satellite off-boresight geometry [24].

$$\hat{b} = -\frac{r_{\text{GPS}}}{\|r_{\text{GPS}}\|}, \tag{19}$$

pointing from the GPS satellite toward the Earth's center. The LNSS lies within the beam if the angle between this boresight vector and the GPS-to-LNSS direction is sufficiently small. The off-boresight angle which captures the geometric alignment of the LNSS with respect to the GPS antenna's coverage cone is computed as:

$$\theta_{\text{off}} = \cos^{-1}\left(\hat{b}^T\,\hat{u}\right) = \cos^{-1}\left(\frac{(-r_{\text{GPS}})^T(r_{\text{LNSS}} - r_{\text{GPS}})}{\|r_{\text{GPS}}\| \cdot \|r_{\text{LNSS}} - r_{\text{GPS}}\|}\right), \tag{20}$$

In addition to the boresight check, the Earth blockage is incorporated to determine whether the LOS vector from the GPS to LNSS intersects the WGS-84 Earth ellipsoid with realistic equatorial and polar radii, respectively. It is implemented using a ray-ellipsoid intersection test in which LOS is excluded if the ray intersects the ellipsoid surface. An atmospheric dip margin is also applied to realize signal degradation due to the Earth's lower atmosphere. This condition accounts for signal attenuation by excluding signals whose path grazes the Earth's atmosphere below a small angular offset of $\varepsilon = 0.5°$ above the Earth's limb. To complement this, the Earth's angular radius as seen from the GPS satellite is computed as:

$$\alpha_{\text{earth}} = \sin^{-1}\left(\frac{R_{\text{earth}}}{\|r_{\text{GPS}}\|}\right), \tag{21}$$

where $R_{\text{earth}}$ is Earth's radius. The GPS-LNSS link is excluded if either $\theta_{\text{off}} > \theta_{\text{cone}}$ or $\theta_{\text{off}} < \alpha_{\text{earth}} + \varepsilon$, reflecting that the LNSS is either outside the transmit beam or behind the Earth from the GPS satellite's viewpoint. This dual-filtering approach enforces realistic Earth-Moon geometry constraints and eliminates physically obstructed signals.

### 2.2.3 Link Budget Constraint Model

If the geometric visibility constraints (i.e., Moon and Earth occultation) are satisfied, an additional communication-related constraint must also be addressed to determine actual signal accessibility. Therefore, a detailed radio frequency link budget model is constructed to evaluate the received signal strength at the LNSS satellite. This model incorporates a realistic GPS Block IIR transmission configuration, with system parameters outlined in Table 1. For consistency, all GPS satellites are assumed to employ identical transmitter gain patterns for Block IIR satellites, as presented in Donaldson et al [24]. This assumption enables uniform application of the gain pattern in the link budget computation while still reflecting realistic system behavior.

The link budget can be analyzed based on Effective Isotropic Radiated Power (EIRP) and $C/N_0$, further details of the link budget design and analysis can be found in [26]. The EIRP from the GPS transmitter is evaluated with the following mathematical equation:

$$EIRP_{\text{dBW}} = P_{\text{tx}} - L_{\text{tx}} + G_{\text{tx}}(\theta_{\text{off}}), \tag{22}$$

where $P_{\text{tx}}$ is the nominal transmit power, $L_{tx}$ represents the system losses, $G_{\text{tx}}$ is the antenna gain, obtained as function of off-boresight angle ($\theta_{\text{off}}$). The signal attenuation due to propagation is computed using the classical Free Space Loss (FSL) formula:

$$FLS_{\text{dB}} = 20log_{10}\left(\frac{4\pi d}{\lambda}\right),\ \lambda = \frac{c}{f} \tag{23}$$

where, $d = \|r_{\text{LOS}}\|$ is the distance between the GPS and LNSS satellites, $c$ is the speed of light, and $f$ is GPS L1 frequency. The carrier-to-noise density ratio at the LNSS receiver is then calculated as:

$$C/N_0 = EIRP_{\text{dBW}} + G_{\text{rx}} - FSL_{\text{dB}} - 10log_{10}(kT), \tag{24}$$

where, $G_{\text{rx}}$ is the LNSS receiver antenna gain, $k$ is Boltzmann's constant, and $T$ is the system noise temperature. The receiver is assumed to have constant gain steerable antenna that perfectly aligns with GPS signal. If the $C/N_0 \geq 15$ dBHz, link between GPS and LNSS is deemed valid. This approach filters out weak or degraded signals, ensuring only practically viable signals would further contribute to position solutions for dilution metrics computation.

**Table 1:** GPS-LNSS link budget parameters.

| Parameter | Symbol | Value |
|---|---|---|
| Transmitter frequency | $f$ | 1575.42 MHz (L1 C/A) |
| Transmitter power | $P_{\text{tx}}$ | 15 dBW |
| Transmitter gain | $G_{\text{tx}}$ | Realistic gain pattern for GPS IIR (13 dBi at 0° off boresight) |
| System losses | $L_{\text{tx}}$ | 2 dB transmission loss, 3 dB (polarization loss) |
| Receiver gain | $G_{\text{rx}}$ | 15 dBi |
| Spectral to Power Density Ratio threshold | $C/N_0$ | 15 dBHz |

### 2.2.4 GPS-LNSS PDOP Computation Model

When at least four GPS satellites meet all visibility and link budget criteria, the geometric quality of the navigation solution is assessed using GPS-LNSS PDOP metric. This involves constructing a geometry matrix using the unit LOS vectors from the LNSS satellite to each visible GPS satellite with the same model as explained in Section 2.1. This PDOP metric captures how the spatial distribution of visible GPS satellites affects positioning errors, with lower values indicating stronger and more favorable GPS-LNSS geometry. The access geometry between GPS system and LNSS satellite differs fundamentally from that of Earth-based users. For lunar satellites, the GPS-LNSS LOS geometry is predominantly more unidirectional, unlike the Earth case where satellites are distributed across the local horizon. This geometric configuration leads to high PDOP along with significant fluctuations in PDOP values over the propagation period. In this study, there is no cut-off imposed for GPS-LNSS PDOP metric computation to assume the worst-case scenario for analysis. The mean PDOP metric is then evaluated over the full dataset using a 2-sigma (95th percentile) filter to exclude extreme outliers and obtain a representative measure of performance.

# 3 Dynamic Model

The dynamic models used in this study consist of two components, i.e. high-fidelity numerical propagation and a $\Delta V$ budget estimation model. The lunar propagator in this study is adopted from [27] and it accounts for perturbation accelerations due to lunar gravity harmonics (degree and order 30), third-body effects from the Earth, Sun, and Jupiter, relativistic corrections, SRP, and Earth albedo. The LNSS effective cross-sectional area and mass are assumed equal to GPS IIR satellites (23 $m^2$, 1080 kg) [28]. The governing equations of motion are integrated using adaptive Runge-Kutta (4,5) method [29]. An output sampling interval (epoch) of 900 s is selected, at which the propagated position and velocity states are recorded for subsequent analysis. These sampled states are then used to compute downstream performance metrics such as lunar observer grid DOPs, GPSS-LNSS line-of-sight availability, and the ΔV budget, etc. Planetary ephemerides and frame transformations have been resolved using NASA's SPICE (Spacecraft Planet Instrument C-matrix Events) toolkit to ensure high precision space geometry and time synchronization [30]. It provides necessary information about ephemerides, reference frames, time conversions, and planetary constants through standardized data. In the propagation model, SPICE functions are employed to extract state vectors and rotation matrices, respectively, ensuring consistency in frame transformations and perturbation acceleration computation.

The numerical integration of the equations of motion is performed in an inertial frame of reference. This frame ensures that the only physical forces dependent accelerations present are considered and prevents fictitious effects arising from frame rotation. We adopt the Moon-Centered Inertial frame aligned with the J2000 equator and equinox (MCI-J2000) as the integration frame. This frame is defined as a non-rotating, right-handed cartesian coordinate system centered at the Moon's center of mass, with its orientation fixed relative to the International Celestial Reference Frame (ICRF) at epoch J2000.0 (i.e., January 1, 2000, 12:00 TT). The Moon-Centered-Moon-Fixed frame (MCMF) has also been exploited to realize lunar rotation for realistic DOP analysis. Similarly, the Earth Centered Inertial frame (ECI-J2000) has also been utilized as a supportive frame for GPS constellation propagation; however, the state matrix of GPS system needs to be transformed from ECI to MCI for LNSS-GPS visibility and DOP analysis.

The Barycentric Dynamical Time (TDB) has been adopted as reference for our lunar propagation model. It is also referred to as Ephemeris Time (ET) and is the uniform time scale used as independent variable in astrodynamics applications throughout the solar system [31]. It represents the time that would be recorded by an ideal atomic clock located at the solar system barycenter, thereby eliminating the irregularities associated with Earth's rotational variations and leap second adjustments. The adoption of TDB ensures temporal uniformity and is particularly important in high-fidelity simulations that include relativistic and third-body perturbations. Moreover, all state vectors and force models are initialized with respect to the standard epoch in J2000. The entire set of propagator parameters are depicted in Table 2.

**Table 2:** Lunar propagator parameters setting.

| Parameter | Setting |
|---|---|
| Propagator | Lunar High Precision Orbit Propagator |
| Integrator | Runge–Kutta (4,5) with AbsTol = $1\times10^{-9}$ and RelTol = $1\times10^{-6}$ |
| Gravity model | Moon_AIUB-GRL350B (30×30) |
| Third body | Sun, Earth and Jupiter as point mass |
| SRP | Mass:1080 kg, Area: 23.78 $m^2$,<br>$CR$ = 1.3 |
| Other perturbations | Relativistic corrections and Earth albedo |
| Reference frames | Inertial MCI and MCMF |
| Propagation time | 10 days with 900-s sampling interval |
| Reference time | Barycentric Dynamical Time |

## 3.1 Station-keeping $\Delta V$ Maneuver Model

The station-keeping $\Delta V$ budget computation is one of the crucial objectives of this LNSS optimization and plays a vital role in selecting a specific LNSS configuration. The $\Delta V$ model is closely linked to the lunar propagator because the LNSS state is derived from the propagation fed to the manoeuvre computation model to compute the orbital drift rate. This drift rate is then processed to extract the thrust components and ultimately $\Delta V$ budget. We have used a highly effective semi-analytical algorithm based on the GVEs to model the station-keeping $\Delta V$ budget. The entire process is described as follows:

### 3.1.1 Orbital Drift Computation

The first step is the extraction of orbital state and determine the time rate of orbital drift under perturbing accelerations. The dynamical state of LNSS constellation configuration comprising of position and velocity vectors in MCI-J2000 frame is converted into classical Keplerian elements at each time step. This transformation is based on Kepler equations derived from two-body orbital mechanics, as detailed in [22]. These elements offer an intuitive description of the orbit and, more importantly, enable direct application of Gauss's variational theory for low-thrust continuous maneuver modelling. Since the numerical integration already incorporates all relevant perturbing accelerations the resulting state vectors inherently reflect the complete dynamical behavior of the system. Therefore, the extracted orbital element sequences capture the full dynamical print of orbital drift in the perturbed environment. The instantaneous secular drift of all the orbital elements is computed by numerical gradient, i.e. first-order time derivatives of each of the orbital elements.

### 3.1.2 Mapping Thrust Components from Orbital Drift

GVEs have been widely applied in astrodynamics to model perturbed orbital motion under disturbing accelerations, as shown in Equations (25)-(29) [32][33]. In the proposed framework, the instantaneous drift of orbital elements is first evaluated using central finite differences between consecutive epochs. The GVEs are then rearranged in inverse form to map the orbital drift into an equivalent continuous-thrust vector expressed in the Radial–Tangential–Normal (RTN) frame, representing the ideal thrust that would fully counteract the perturbation-induced drift at that epoch. It is important to emphasize that this methodology is used solely to quantify the $\Delta V$ budget for station-keeping by capturing all short-period perturbation effects and nonlinearities that are often neglected in analytical approaches based on mean orbital elements or secular perturbation theory. This approach results in more accurate and high-fidelity station-keeping $\Delta V$ estimates as it is based on real-time propagated state. However, the framework should not be interpreted as a maneuver implementation strategy. In practice, station-keeping is carried out through finite thrusting, applied at selected orbital locations (e.g., near perilune or apolune) once deviations exceed predefined thresholds. The continuous-thrust representation is therefore a modelling tool for $\Delta V$ estimation rather than a prescription of how maneuvers would be executed operationally. Physically, the thrust vector can be resolved into three components in RTN, i.e. $T_R$, $T_T$, and $T_N$, where $T_N$ rotates the orbital plane by changing inclination and RAAN, $T_T$ modulates the orbital energy for SMA control and $T_R$, couples into eccentricity and AOP corrections.

$$\dot{a} = \frac{2a^2}{h}\left(T_R e \sin v + T_T \frac{p}{r}\right), \tag{25}$$

$$\dot{e} = \frac{1}{h}\left(T_R p \sin v + T_T (p + r) \cos v + re\right), \tag{26}$$

$$i = T_N \frac{r \cos \theta}{h}, \tag{27}$$

$$\dot{\Omega} = T_N \frac{r \sin \theta}{h \sin i}, \tag{28}$$

$$\dot{\omega} = \frac{1}{he}\left(-T_R p \cos v + T_T (p + r) \sin v\right) - T_N \frac{r \sin \theta \cos i}{h \sin i}, \tag{29}$$

In the above equations, $a$ is SMA, $e$ is Eccentricity, $i$ is Inclination, $\Omega$ is RAAN, $\omega$ is AOP, $v$ is true anomaly, $p$ is semi-latus rectum, h is specific angular momentum magnitude, $r$ is orbital radius at a specific true anomaly, θ is argument of latitude and $[\dot{a}, \dot{e}, i, \dot{\Omega}, \dot{\omega}]$ represent the time derivatives of the Keplerian elements.

Notably, we do not correct true-anomaly drift because it evolves continuously under Kepler's laws and cannot be frozen. Likewise, when initializing our thrust-allocation model we assume the radial contribution of the SMA drift to be zero for elliptical orbits and assign the entire thrust acceleration to be purely tangential. This is because, even for highly eccentric lunar orbits, tangential acceleration remains the main contributor of SMA variations. Therefore, this simplification captures the dominant component of required thrust acceleration for SMA corrections. Any smaller radial contribution that does arise can then be applied later as a modest corrective impulse (if needed). Finally, to avoid numerical singularities at small inclination or eccentricity, each individual acceleration component is capped so that the solution remains stable for near-circular equatorial orbits. The thrust acceleration components are derived as follows:

$$T_{SMA} = \frac{hr}{2pa^2}\dot{a}, \qquad T_T = T_{SMA} \tag{30}$$

$$T_i = \frac{h}{r\cos\theta}i, \qquad T_\Omega = \frac{h\sin i}{r\sin\theta}\dot{\Omega}, \tag{31}$$

$$T_N = \sqrt{{T_i}^2 + {T_\Omega}^2}.\cos\psi_N \tag{32}$$

$$T_e = \frac{h\dot{e} - T_T\,[(p+r)\cos v + re]}{p\sin v}, \tag{33}$$

$$T_\omega = \frac{1}{p}\left[-he\left(\dot{\omega} + \frac{T_N r\sin\theta\cos i}{h\sin i}\right) + T_T(p+r)\sin v\right]\cos v, \tag{34}$$

$$T_R = T_e + T_\omega \tag{35}$$

The inclination and RAAN corrections require accelerations along two orthogonal directions within the common normal (out-of-plane) axis. Instead of treating these quantities as scalars, we preserve the directional component of the out-of-plane plane thrust using $\psi_N = \tan^{-1}\left(\frac{T_\Omega}{T_i}\right)$. On the other hand, the eccentricity and AOP corrections lie along the same radial direction, allowing their scalar magnitudes to be added directly to compute the total radial thrust component. The total instantaneous thrust acceleration required to freeze the orbital drift is given by:

$$T_{\text{total}}(t) = \sqrt{T_R^2(t) + T_T^2(t) + T_N^2(t)} \tag{36}$$

### 3.1.3 Station-keeping $\Delta V$ Computation and Optimization

The magnitude of the required thrust vector, $T_{\text{total}}(t)$ defines the instantaneous acceleration expenditure needed to freeze the element set at their initial levels. Integrating this magnitude over the full propagation interval $[t_0, t_f]$ yields the $\Delta V$ budget, as expressed in Equation (37). Because the thrust directions vary continuously with the natural motion of the satellite, trapezoidal numerical integration is preferred for its second-order accuracy and simplicity. This baseline $\Delta V$ can be interpreted as an upper bound which assumes that spacecraft can accurately produce the exact control acceleration completely mitigate the secular orbital drift due to numerous perturbations at a specific instant irrespective of the energy-optimal steering.

$$\Delta V = \int_{t0}^{tf} T_{\text{total}}(t)dt \tag{37}$$

In practice, spacecraft are subject to design limitations such as finite thrust levels, duty cycles, and maneuver execution constraints. To address this, we extend the $\Delta V$ estimation with a nonlinear optimization framework that allows for adjustments in thrust orientation when the exact required magnitude is not available. By slightly steering the direction of the thrust vector in the local RTN frame, we can often achieve the same component ratios *(*$T_R$*,* $T_T$ and $T_N$*)* while cutting off a small amount of thrust magnitude. In other words, if the original thrust components lie along a specific direction in RTN frame, there might exist a slightly different orientation that would still be colinear to the required thrust angles but uses lower thrust.

To implement this concept at a discrete time $t_{(k)}$, we have introduced three decision variables, i.e. the actual thrust magnitude $T_{\max} \geq 0$ along with two thrust angles: azimuth in the orbital plane *(α)*, measured from the radial axis toward the tangential axis and elevation *(β)* above the orbital plane into the normal direction [34]. If thrust of magnitude $T_{\max}$ at a specific thrust angles is applied, its RTN components can be derived from:

$$\begin{cases} T_R = T_{\max} \cos\beta \cos\alpha \\ T_T = T_{\max} \cos\beta \sin\alpha \\ T_N = T_{\max} \sin\beta \end{cases} \tag{38}$$

We have applied an adaptive weighting approach in which the total (original) components determine the proportion of total thrust allocated in each direction relative to the required threshold. If we normalize the original thrust vector components into weights, it will compel the new thrust angles to be parallel with the original ones with lower thrust magnitude. However, this thrust vector still addresses orbital drift in the exact same proportion as the original computation but with lower thrust magnitude.

$$w_R = \frac{|T_R|}{|T_R| + |T_T| + |T_N|}, w_T = \frac{|T_T|}{|T_R| + |T_T| + |T_N|}, w_N = \frac{|T_N|}{|T_R| + |T_T| + |T_N|} \tag{39}$$

$$\begin{cases} w_R|T_R| = T_{\max} \cos\beta \cos\alpha \\ w_T|T_T| = T_{\max} \cos\beta \sin\alpha \\ w_N|T_N| = T_{\max} \sin\beta \end{cases} \tag{40}$$

where $w_T$ is the adaptive weight for tangential thrust magnitude, $w_N$ is the adaptive weight for normal thrust magnitude and $w_N$ is the adaptive weight for radial thrust magnitude. We then solve small-scale Non-Linear Programming (NLP) at each time step to minimize $T_{\max}$ magnitude by steering the thrust angles within the defined bounds, i.e. $\alpha \in [-\pi, \pi]$ and $\beta \in [-\frac{\pi}{2}, \frac{\pi}{2}]$. The original thrust magnitude $T_{\text{total}}$ and corresponding thrust angles are taken as initial guess. The NLP objective function is mathematically represented as follows:

$$\min_{(T_{\max},\ \alpha,\beta)} = \{T_{\max} + \lambda(T_{\max} - T_{\text{total}})^2\}, \qquad 0 \leq T_{\max} \leq T_{\text{total}}, \alpha \in [-\pi, \pi], \qquad \beta \in \left[-\frac{\pi}{2}, \frac{\pi}{2}\right] \tag{41}$$

where the penalty weight $\lambda = 20$ discourages excessive deviation of $T_{\max}$ from $T_{\text{total}}$ . We have solved this three-variable based constraint problem via Sequential Quadratic Programming (SQP) [35]. The solver iterates until the Karush–Kuhn–Tucker conditions are satisfied to produce optimized $T_{\max}$ along with optimum thrust angles. This optimized thrust can then be integrated over time to compute optimized station-keeping $\Delta V$. Importantly, this optimization step is mainly included to demonstrate how thrust vector steering can achieve adequate orbital element corrections even under limited thrust availability. In this sense, this small-scale NLP optimization not only provides a theoretical minimum executable $\Delta V$ estimates but also a framework to the mission planners to handle the practical constraints of maneuvers during implementation phase.

# 4 Study Limitations:

There are some limitations in the computation of performance evaluation and $\Delta V$ estimation model in this study. The performance of LNSS constellation is evaluated solely by using the PDOP and HDOP metrics. These are purely geometric measures of satellite-receiver geometry and do not account for time-related errors such as receiver clock offsets, biases, or synchronization errors. Consequently, while the PDOP and HDOP based analysis provides an accurate assessment of the geometric positioning quality, it does not capture potential degradations due to clock errors or other system biases that would influence the overall navigation solution, such as reflected in GDOP.

The station-keeping $\Delta V$ budget is expressed in $\text{kms}^{-1}$ per satellite per year, extrapolated from the 10-day simulation results. This approach is adopted because performing a full one-year simulation would be computationally expensive. Moreover, the framework assumes continuous thrust applied at every epoch. While this approach allows for high-fidelity quantification of the station-keeping $\Delta V$ under idealized and optimized conditions, it is not directly representative of practical maneuver execution, where thrust is applied in finite impulses or limited-duration arcs at selected orbital positions. The model also does not account for maneuvers execution errors, such as misalignment of thrust vectors, uncertainties in thrust magnitude, or propellant limitations. Therefore, the station-keeping $\Delta V$ budget estimates obtained here should be interpreted as idealized requirements that quantify the effort needed to maintain the orbit in perturbated cislunar environment. These estimates serve as a benchmark for constellation optimization, rather than precise mission planning values, and provide a framework to guide practical maneuver strategies under real-world constraints.

# 5 Optimization Framework

The optimization of LNSS is inherently a multi-objective problem involving conflicting design trade-offs between geometric performance, coverage availability, integrity, and orbital station-keeping requirements. To efficiently navigate this high-dimensional design space, we employed NSGA-II in MATLAB, which is a widely adopted evolutionary algorithm particularly suited for handling complex, non-linear, and multi-modal objective landscapes without requiring gradient information [36]. The core principle of NSGA-II is to evolve a population of candidate solutions across generations using genetic operators like selection, crossover, and mutation, etc. while maintaining a diverse set of non-dominated solutions that approximate Pareto-optimal front. The algorithm performs non-dominated sorting to rank individuals based on Pareto dominance. If $f_i$ and $f_j$ are two objective functions, a solution $x_1$ dominates another $x_2$ if $f_i(x_1) \leq f_i(x_2)$ for all objectives $i$ and if $f_j(x_1) < f_j(x_2)$ for at least one $j$. To preserve diversity along the front, it computes the crowding distance $d_i$ for each solution in a Pareto front, defined as the average normalized distance to its two nearest neighbors in objective space. Solutions with larger $d_i$ values are preferred during selection, promoting uniform front distribution [37]. Genetic operators govern evolutionary dynamics and improve the solution until user defined convergence is achieved. The optimization terminates upon reaching a maximum number of generations or a specified tolerance threshold. The objective function of this problem can be categorized into two main parts including the design variables and objectives that are explained as follows:

## 5.1 Design Variables

The design variables include orbital elements (Keplerian) essential to define geometry of the LNSS constellation in MCI frame. These design elements need to be initialized based on some bounds as depicted in Table 3. These bounds have been selected based on mission design requirements; for instance, the SMA lower bound is selected to ensure that the lunar orbits are stable for longer operations and the upper bound corresponds to approximately Earth GPS system altitude conditions. This altitude would produce very similar communication constraints as of the GPS system. Similarly, the eccentricity bounds ensure diversity ranging from circular to highly elliptical obits, leveraging the regional / customized coverage, e.g. lunar south pole. The inclination is assigned a full search space between the equatorial and the polar orbits and AOP has also been given full range of flexibility in search space. This flexibility would also be useful for the analysis of $\Delta V$ dependency on the orbital configuration of the solutions. The total number of satellites and planes are bounded by considering the mission design cost factors, i.e. reducing the total number of satellites, while achieving mission design objective. When the algorithm initializes, it selects a combination of the design / decision variables and loops over each plane index, $p = 1, \dots N_\mathrm{p}$ to form walker configuration. The RAAN of $p^{th}$ orbital plane can mathematically be written as [34]:

$$\Omega_p = (p-1)\frac{2\pi}{N_\mathrm{p}}, \tag{42}$$

where $N_\mathrm{p}$ is the number of orbital planes. It implies that the LNSS planes of each of the constellation configurations have been uniformly distributed in 360° of RAAN spread with $N_s$ (number of satellites) also uniformly spaced by mean anomaly in each of the orbital plane. In other words, the constellation design follows the Walker configuration to ensure design symmetry for convenient $\Delta V$ modelling.

When using population-based methods like NSGA-II, maintaining a normalized and uniformly distributed design space is essential for efficient convergence. For this purpose, all design variables are scaled to unitless quantities before being passed to the objective function. This normalization procedure enables the genetic algorithm to operate effectively within a bounded and homogeneous search domain, independent of the underlying physical units or magnitudes of the parameters. As we know the set of design variables is $X = [\mathrm{a}, \mathrm{e}, \omega\ \mathrm{i}, N_p, N_s]$ we define a set of corresponding step sizes $\Delta X = [\Delta a, \Delta e, \Delta\omega\ \Delta i, \Delta N_p, \Delta N_s]$ to scale the variables to a unitless domain for optimization. The scaled representation $x = [x_1, x_2, \dots, x_6]$ is computed using:

$$x_j = \frac{X_j - L_j}{\Delta X_j}, \mathrm{j} = 1,2 \dots, 6, \tag{43}$$

where $L_j$ is the lower bound of variable $X_j$. This transformation maps each parameter to an integer-like scaled space, ensuring that the search grid aligns with user-defined discretization levels. After the optimization algorithm proposes new scaled values x, the original dimensional values of X are reconstructed (de-scaled) using:

$$X_j = \min\left(\max\left(L_j + \Delta x_j\,.\,X_j, L_j\right), U_j\right), \tag{44}$$

where $U_j$, is the upper bound of the variable. The integer-valued variables such as $N_p$ and $N_s$ are rounded to the nearest integer to ensure feasible satellite constellation configurations.

## 5.2 Fitness variables

The fitness (objective) variables include DOP matrices for lunar south pole users, PDOP metric for GPS and LNSS integration and station keeping $\Delta V$ budget. These metrics have already been thoroughly explained in Section 2 and 3. Like design variables, the mean values of fitness variables are also bound by constraints as depicted in Table 3. The main purpose of these benchmarks is to guide the algorithm to explore only high performing solutions. In other words, it would allow the optimizer to search for the non-dominating solution in those regions where the objectives fulfill the navigation performance and $\Delta V$ budget requirements.

The optimizer identifies LNSS solutions with best trade-offs in each of the iterations. For decision making, these candidate solutions are represented by a vector in six-dimensional design space, e.g. $x \in R^6$ and the mean values fitness variables. The objective function of LNSS optimization can be expressed by the following equation:

$$[N_{sat}, \overline{PDOP_{3\sigma}}, -PDOP_{avail}, \overline{HDOP_{3\sigma}}, -HDOP_{avail}, \Delta V_{total}, \Delta V_{opt}, -\overline{VIS_{GPS-LNSS}}, \overline{PDOP_{GPS-LNSS}}\ ] = \boldsymbol{f}(a, e, \omega\ i, N_p, N_s), \quad (45)$$

where $N_{\text{sat}}$ is the total number of satellites in the constellation, $\overline{PDOP}$ and $\overline{HDOP}$ are the mean 3-sigma dilutions for south pole surface users, $-PDOP_{\text{avail}}$ and $-HDOP_{\text{avail}}$ represent the percentage of navigation coverage availability for the users (negative sign indicate that these objectives are to be maximised), $-\overline{VIS_{\text{GPS-LNSS}}}$ is the mean GPS satellite visibilty for LNSS satellite, $PDOP_{\text{GPS-LNSS}}$ is the mean 2-sigma PDOP of GPS system for the LNSS satellite, $\Delta V_{\text{opt}}$ and $\Delta V_{\text{total}}$ are the optimized and total station-keeping $\Delta V$ mean values, respectively.

**Table 3:** Design variable and fitness variable bounds.

| | | Type | Lower bound | Upper bound |
|---|---|---|---|---|
| **Design Variables** | $a$ | Real | 4000 km | 24000 km |
| | $e$ | Real | 0 | 0.8 |
| | $\Omega$ | Real | 0° | 359° |
| | $i$ | Real | 1° | 90° |
| | $\omega$ | Real | 1° | 359° |
| | $N_p$ | Integer | 2 | 5 |
| | $N_s$ | Integer | 1 | 4 |
| **Fitness Variables** | $N_{sat}$ | Integer | 1 | 20 |
| | $PDOP_{3\sigma}$ | Real | < 0 | 10 |
| | $PDOP_{avail}$ | Real | 70 | 100 |
| | $\Delta V_{total}$ | Real | 0 kms$^{-1}$ per year | 1 kms$^{-1}$ per year |

## 5.3 Implementation logic

The NSGA-II optimizer is integrated with the lunar propagator, navigation performance evaluation model and $\Delta V$ model based on modular scripts in MATLAB. The simulation duration is 10 days starting at 2025 May 01 00:00:00 ET to 2025 May 11 00:00:00 ET with 900-s sampling interval. The implementation logic of optimization algorithm is illustrated in Figure 2 and explained as follows:

- ***Initialization:*** The LNSS constellation is initialized based on the design variable bounds as described in Table 3. Similarly, GPS constellation is also initialized from the TLE data. The performance metrics, observer grid and propagation are integrated within the optimization framework.
- ***Propagation:*** The dynamic models propagate LNSS orbital configuration and GPS constellation for the simulation time of 10 days. The LNSS is propagated with predefined high fidelity numerical lunar propagator and GPS system is propagated in two-body system for simplicity. The state (position and velocity) of both systems has been sampled in respective inertial frames at specific epochs every 900-s interval. This data is required to compute DOP metrics for lunar user gird and PDOP metrics for GPS-LNSS integrity.

  ***DOP metrics computation:*** The PDOP and HDOP metrics are computed for the receiver grid within 60°S to 90°S in MCMF at each time epoch. This calculation is based on LOS unit vectors from all grid points to all LNSS visible satellites with 5° mask angle constraint and eventually construct the geometry matrix (if 4-satellites are visible) and then dilutions. The instantaneous data of DOP metrics is used to compute mean values to be incorporated into objective function for decision making process.

For GPS to LNSS PDOP analysis, the LOS vector between GPS system and first satellite of each LNSS configuration is computed in MCI frame. This computation is subject to lunar and Earth geometric occultation and communication constraints (EIRP, FSL, $C/N0$ dB-Hz, etc.). The LOS between GPS and the LNSS satellite are only considered valid if they are occluded from the Earth and the Moon and $C/N0$ is 15 dB-Hz. If at least four GPS satellites are accessed by the LNSS satellite, the LNSS-GPS geometry matrix is constructed to calculate PDOP matric.

- ***Station-keeping $\Delta V$ computation:*** The instantaneous orbital elements is computed for first satellite of each LNSS configuration from the state propagation data. This orbital data is used in drift rate computation of each element at any given epoch. The sequence then computes the required thrust accelerations with inverted GVEs and integrating total thrust magnitude is to get total $\Delta V$ budget estimation.
  To implement $\Delta V$ budget optimization, the thrust vector ($\alpha$ *and* $\beta$) is computed using thrust components in RTN frame using equation (38). A small-scale NLP is solved for each epoch to optimize thrust vector and thrust magnitude to compute optimized thrust components. This optimized thrust magnitude $T_{opt}(t)$ is integrated over time to generate optimized $\Delta V$ estimation.
- ***Configuration update and Pareto search:*** All final values including mean DOPs, DOP availability and $\Delta V$ estimates per year are incorporated into the fitness function and the LNSS configuration is updated. If the values of fitness variables are within the defined bounds, the corresponding LNSS solution is selected for the analysis of trade-offs between objectives as defined in equation (42). The optimizer then identifies Pareto-optimal / the non-dominant solutions in each iteration.

It must be noted that station-keeping $\Delta V$ computations and GPS–LNSS PDOP metric are computed only for the first satellite of each LNSS configuration. This simplification reduces execution time of the framework and is justified because the Walker constellation exhibits symmetrical orbital geometry, meaning that all satellites experience nearly identical $\Delta V$ requirements and GPS visibility characteristics.

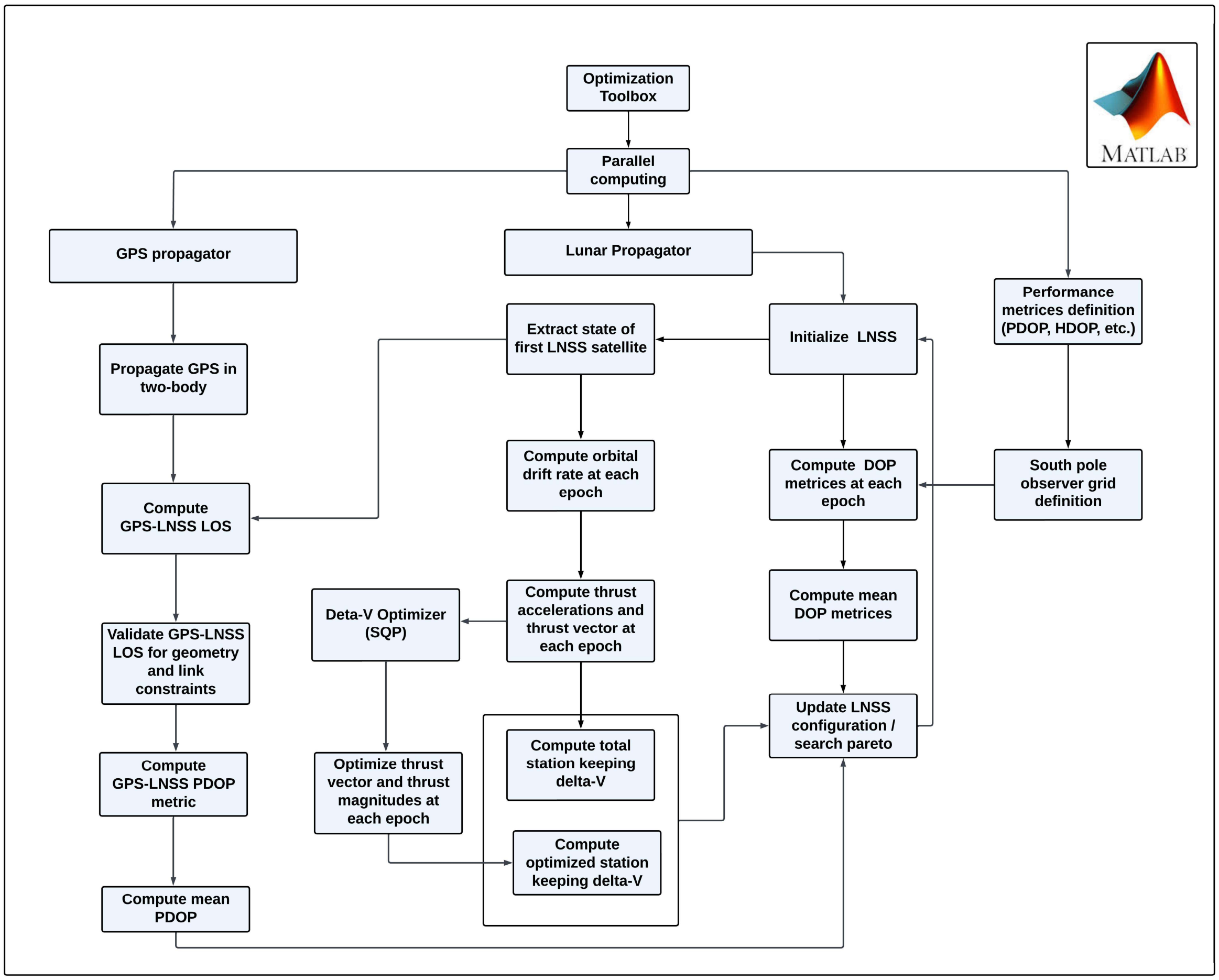

**Figure 2:** Illustration of simulation setup for complete optimization framework.

# 6 Results and Analyses

We have conducted two simulation studies to demonstrate the working mechanism of our methodology. First simulation focuses on real-time evaluation of the objective functions (e.g., performance metrics) within a standalone LNSS constellation. This simulation is designed to explicitly illustrate the step-by-step process through which objective values are derived from the constellation geometry and dynamics. Notably, this preliminary analysis does not involve the NSGA-II optimization framework; rather, it serves to understand the formulation and computation of the performance metrics used in the optimization. In contrast, the second simulation integrates the full LNSS optimization framework as described in Section 5 where the NSGA-II algorithm is employed to explore appropriate constellation designs.

## 6.1 Standalone Configuration Simulation

We modelled an LNSS configuration based on an ELFO-type constellation architecture, similar to one of the designs proposed in [9]. The orbital elements of this LNSS constellation model are mentioned in Table 4 and visualized in MCI frame in Figure 3. The constellation is propagated using our high-fidelity lunar orbit propagation model to demonstrate and understand the objective function computation mechanism. It must be noted that these metrics have been computed without optimization and results demonstrate and visualize the ability to of our simulation to efficiently calculate objective function variables.

The dilutions and coverage are calculated for the observer grid on the lunar south pole (already explained) and the mean values are given in Table 4. The temporal variation of the mean satellite visibility and the dilutions of the spatiotemporal observer grid for the entire simulation duration are depicted in Figure 4. The PDOP and HDOP are excellent and stable for the entire simulation span due to uniform high satellite visibility. The progression of station-keeping $\Delta V$ with orbital evolution of the first satellite of the constellation and thrust acceleration have been analyzed in Figure 5 and Figure 6. The drift of orbital elements because of perturbation forces for a span of 10 days is depicted in Figure 5. This instantaneous orbital evolution data (differential gradient of orbital elements) is further processed to compute total and optimized normal, tangential and radial thrust acceleration magnitudes as shown in Figure 6. These thrust magnitudes are derived by incorporating orbital drift rates in the GVEs. As the orbits of this system a highly eccentric, the altitude of satellite continuously varies during the orbital motion resulting in non-uniform orbital drift rates throughout the propagation. For instance, we can observe numerous spikes in thrust acceleration graphs for the entire course simulation. These spikes correspond to the rapid instantaneous fluctuation of the relevant orbital elements at the perilune (closest point). The reason that these orbital oscillations exhibit pronounced spikes at perilune because gravity-harmonic perturbations grow rapidly with decreasing altitude and thus are strongest when the satellite is at its minimum orbital radius. These thrust components are then used to compute the total and optimized $\Delta V$ as given in Table 4. In a similar manner, the GPS-LNSS visibility and PDOP have also been computed for this standalone LNSS model. The GPS constellation is propagated with a two-body model in ECI-J2000 frame. The instantaneous visibility and PDOP of this analysis are depicted in Figure 7. We can clearly see that there are some instances when GPS-LNSS visibility becomes zero due to the geometric and/or link constraints. As GPS-LNSS link cannot be established, thus PDOP plot also shows blank spots during these instances. The mean GPS-LNSS PDOP for the simulation span is 6238 as depicted in Table 4. Since this metric heavily relies on LNSS orbital arrangement, we plan to identify LNSS configurations with better GPS-LNSS PDOP mean value in our optimization.

**Table 4:** Standalone LNSS constellation orbital elements and mean performance indicators.

| | Parameter | Value |
|---|---|---|
| **Orbital configuration** | SMA | 6143 km |
| | Eccentricity | 0.6 |
| | Inclination | 51.7° |
| | AOP | 90° |
| | No of Planes | 2 (Ω = 0°, 180°) |
| | No of Satellites | 16 |
| **Performance metrics** | Satellite visibility | 11.22 |
| | $\overline{PDOP_{3\sigma}}$ and $\overline{HDOP_{3\sigma}}$ | 2.17, 0.83 |
| | $PDOP_{\text{avail}}$ and $HDOP_{\text{avail}}$ | 100%, 100% |
| | $\Delta V_{\text{total}}$ per satellite | 0.50 $kms^{-1}$ / year |
| | $\Delta V_{\text{opt}}$ per satellite | 0.37 $kms^{-1}$ / year |
| | $\overline{VIS_{\text{GPS-LNSS}}}$ | 7.74 |
| | $PDOP_{\text{GPS-LNSS}}$ | 6238 |

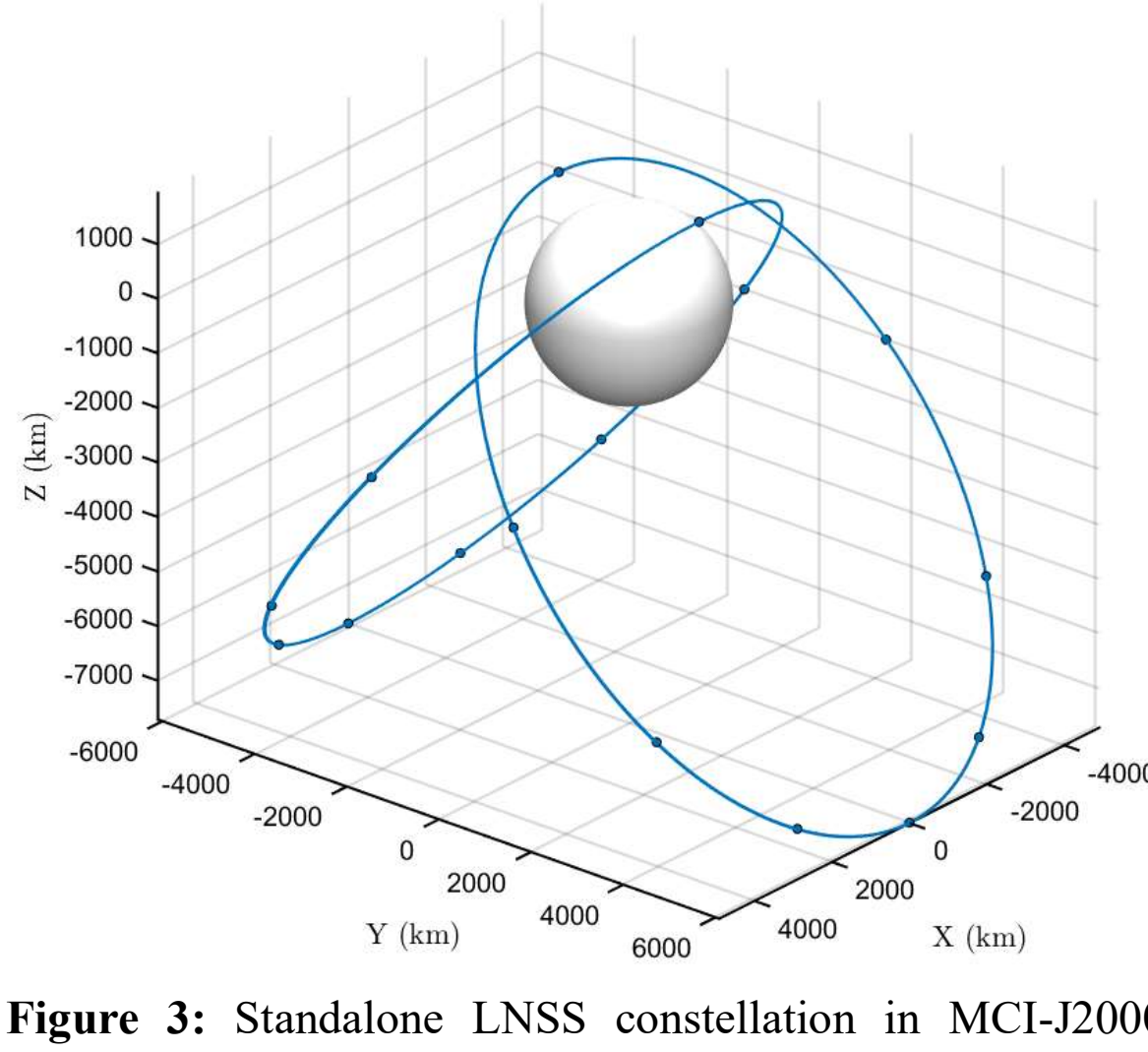

**Figure 3:** Standalone LNSS constellation in MCI-J2000 frame.

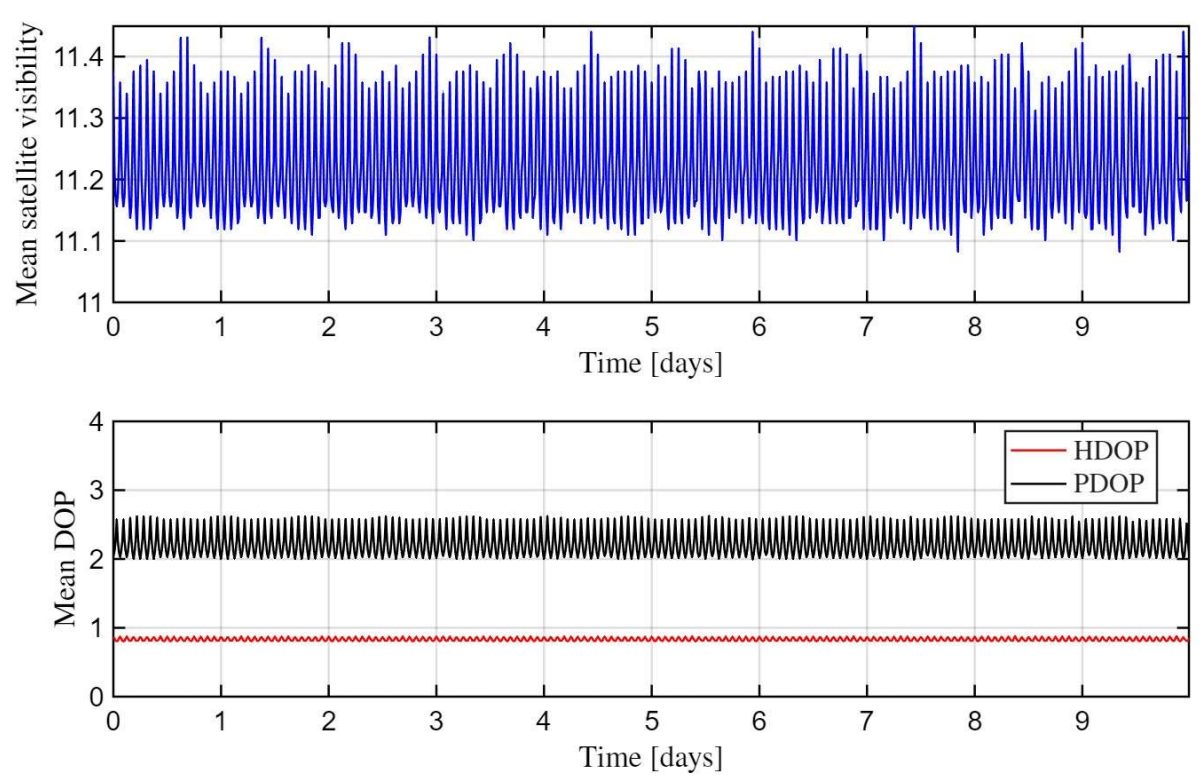

**Figure 4:** Temporal variation in mean visibility and dilutions for the observer grid on south pole.

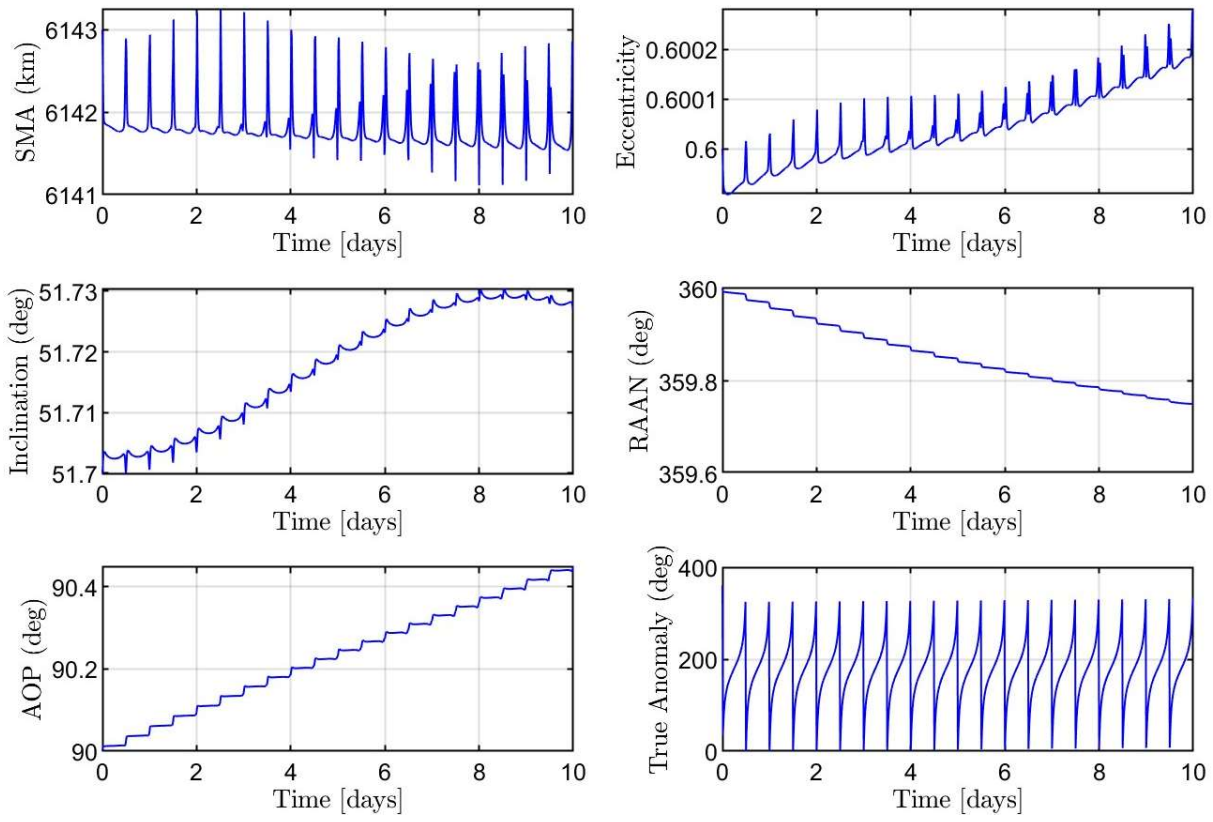

**Figure 5:** Evolution of orbital elements of the standalone LNSS first satellite under the influence of perturbations in the MCI-J2000 frame.

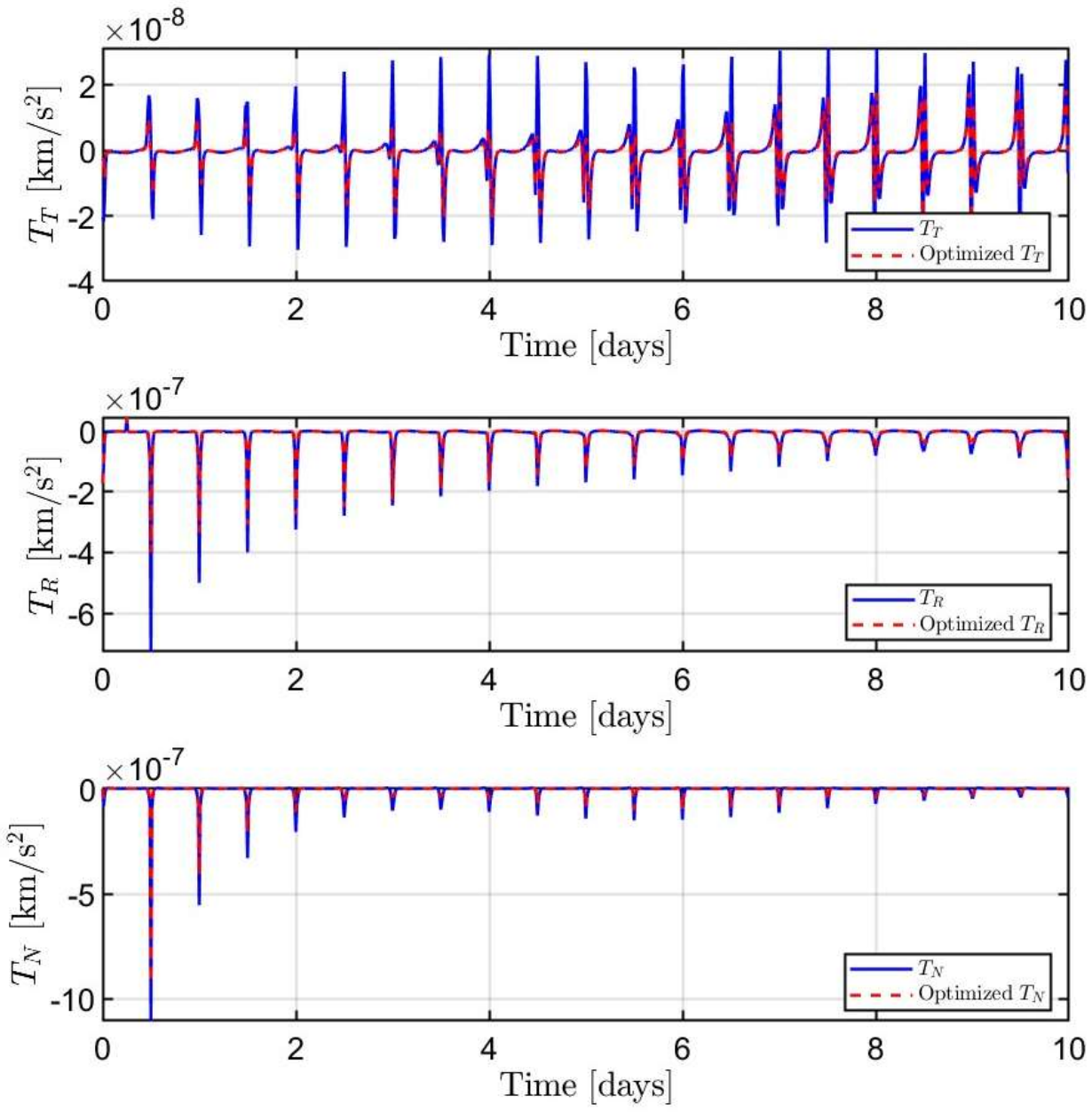

**Figure 6:** Instantaneous thrust magnitude (Total and optimized) for the first satellite of standalone LNSS constellation.

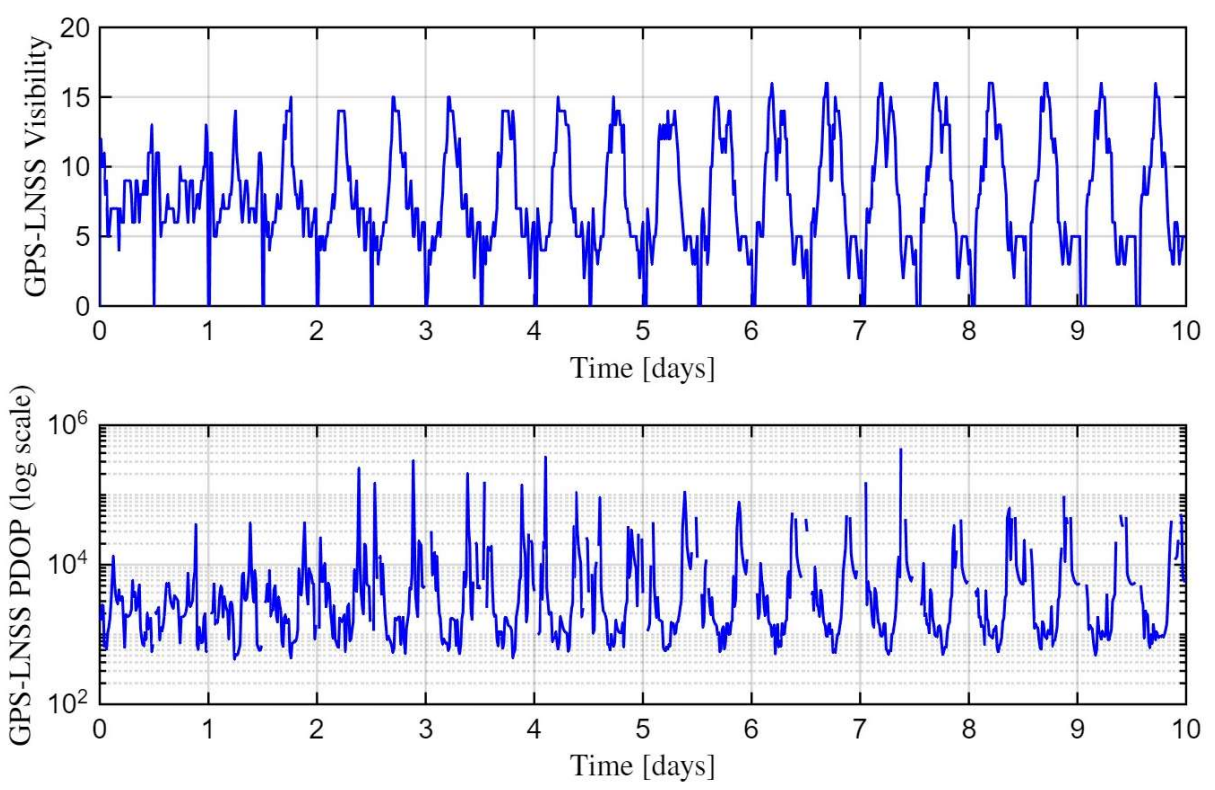

**Figure 7:** GPS-LNSS standalone instantaneous visibility and PDOP in logarithmic scale.

## 6.2 MOGA Simulation

In this section, fully integrated optimization framework setup has been explained in detail which includes the setup parameters, results and algorithm evaluation. To effectively explore the high-dimensional and multi-modal design space of the LNSS optimization problem, a balanced combination of NSGA-II parameters has been employed within MATLAB environment to get diverse yet optimal solutions. The population size is selected as 120 to ensure sufficient genetic diversity across the six-dimensional decision space, enabling a broad search that mitigates premature convergence. This value strikes a balance between solution diversity and computational feasibility, especially considering the high evaluation cost of the fitness function involving orbital propagation, GPS–LNSS visibility analysis, and $\Delta V$ computation. The initial population is selected with a fixed random seed, enabling consistent initial population generation and stochastic operations in the genetic algorithm. The algorithm is run for 1000 generations, which was empirically determined to be adequate for the convergence of the Pareto front while maintaining tractable runtime. A crossover fraction of 0.65 was adopted to ensure that offspring inherit traits from both parents in a balanced manner, promoting a healthy trade-off between exploration and exploitation. Meanwhile, to handle the mixed-integer nature of the decision variables, adaptive mutation function is employed within the defined variable bounds. The Pareto fraction is chosen as 0.3, allowing only highly elite non-dominated solutions to survive into the next generation, thereby preserving diversity along the front. A function tolerance of $1\times10^{-6}$ has been imposed to halt the search when improvements in the objective values became negligible. This high tolerance is selected to prevent premature convergence on fewer iterations. These operational parameters of NSGA-II algorithm are tabulated in Table 5.

**Table 5:** NSGA-II optimizer parameters setting.

| Parameter | Value |
|---|---|
| Population size | 120 |
| No. of generations | 1000 |
| Total Evaluations | 120,000 |
| Crossover fraction | 0.65 |
| Mutation function | Adaptive mutation |
| Pareto fraction | 0.30 |
| Function tolerance | $1\times10^{-6}$ |

The optimizer has identified 5622 unique LNSS constellation configurations that fulfill the lunar navigation positioning and integrity requirements in a constraint setting. The results showing the best trade-offs between the key conflicting objective function variables have been shown in Figure 8 and Figure 9. The Pareto front of the performance metrics (PDOP and HDOP), station-keeping $\Delta V$ and mission design cost (number of satellites) is presented in Figure 8. It must be noted that minimum executable $\Delta V$ or $\Delta V_{\text{opt}}$ is considered for Pareto fronts generated for analysis of optimization results. The best trade-offs of these objectives lie within the boundary of densely populated region at the bottom of the scatter plot, e.g. having $\Delta V$ less than 0.4 kms-1 per satellite per year and for a range of PDOP and HDOP values and satellite number. Similarly, Figure 9 illustrates the trade-offs in 4-objectives space between the LNSS navigation performance (PDOP), GPS-LNSS integrity (GPS-LNSS PDOP), $\Delta V$ budget and total satellites. It must be noted that the GPS-LNSS PDOP is independent of any bound in objective function to fully visualize its relationship LNSS orbital geometry (discussed later). The results suggest that GPS-LNSS PDOP can range within 3862-10883 for numerous Pareto optimal solutions. For Earth GPS integrity perspective, the optimum LNSS configuration lies within the region having $\Delta V$ less than 0.4 kms$^{-1}$ per satellite per year and PDOP around 4000 - 5000.

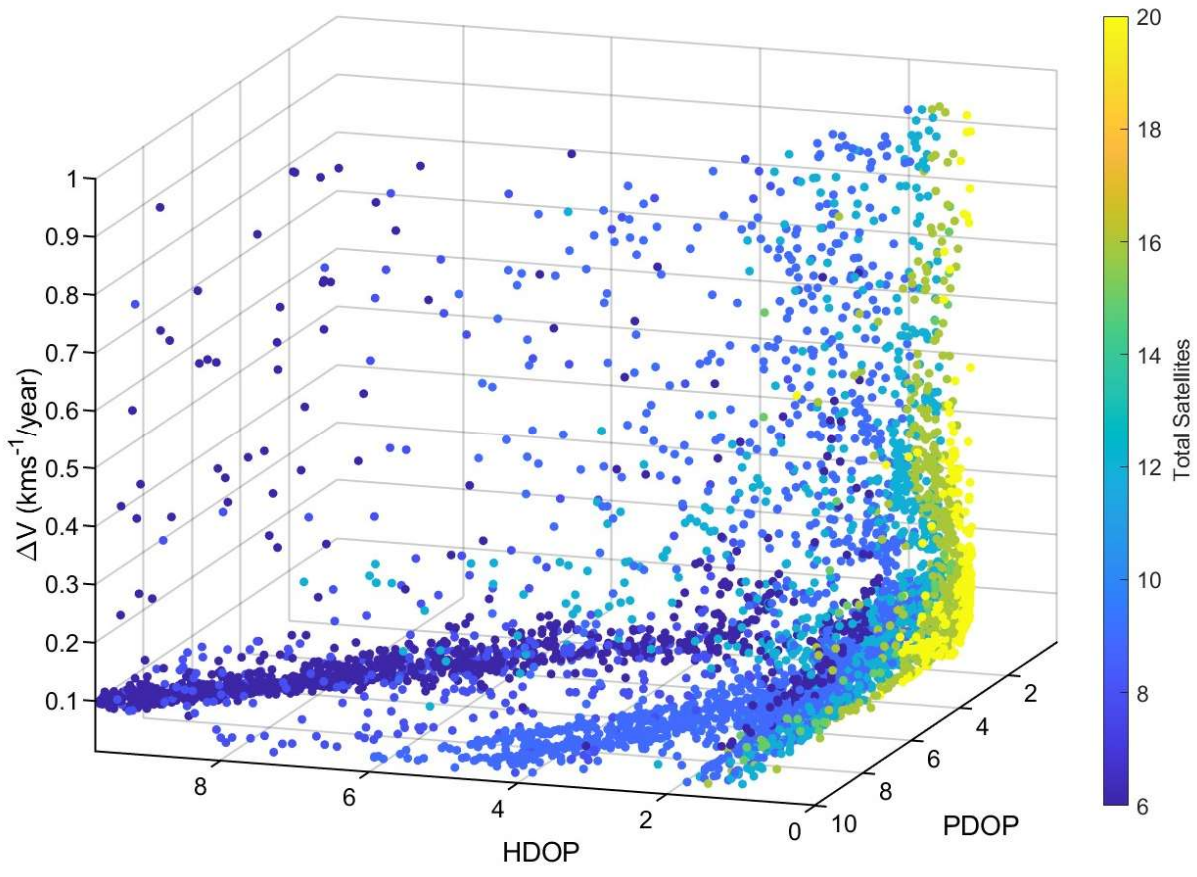

**Figure 8:** LNSS configurations in four-objective space showing trade-off between HDOP, PDOP and station-keeping $\Delta V$ in km$^{-1}$ per satellite per year (Total satellites as color-coded).

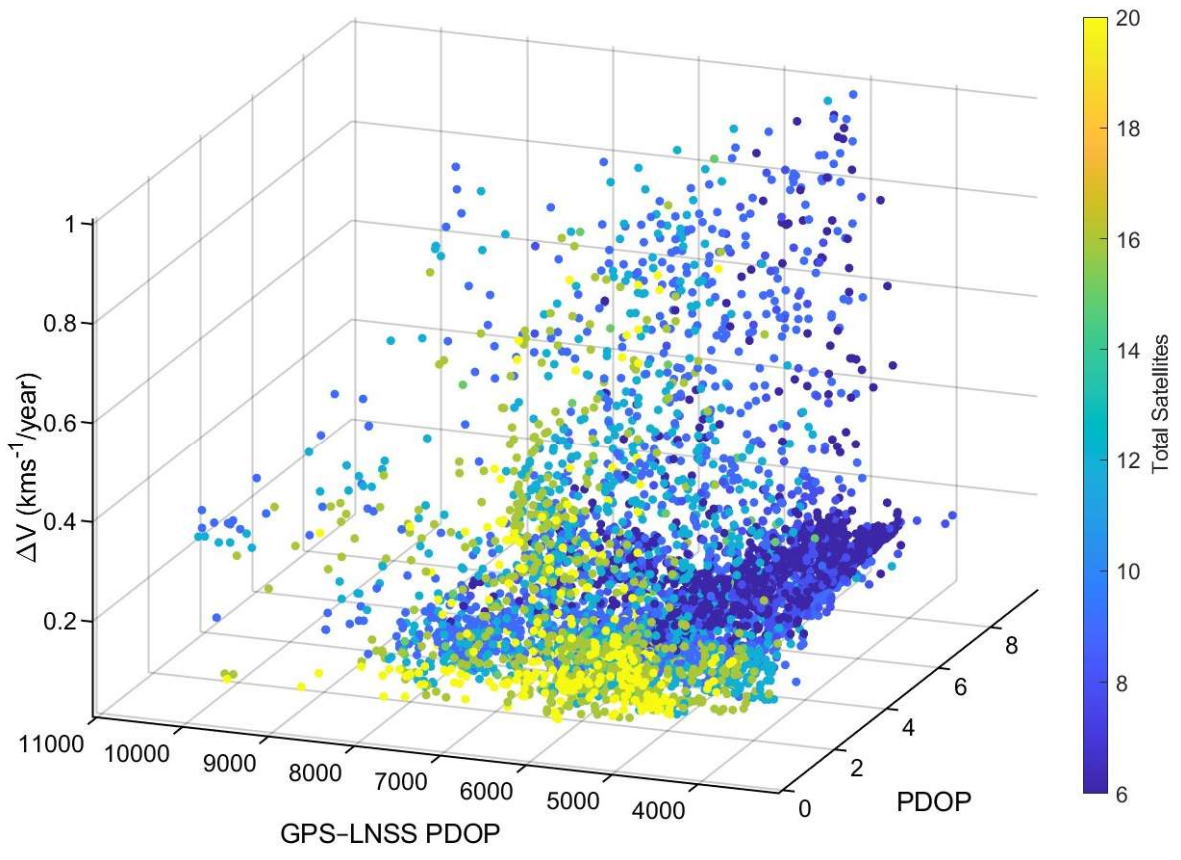

**Figure 9:** LNSS configurations in four-objective space showing trade-off between PDOP, station-keeping $\Delta V$ (km$^{-1}$ per satellite per year) and GPS-LNSS PDOP (Total satellites as color-coded)

The solutions that lie on the left bottom of Figure 10 represent moderate PDOP (6-10) with 70-80 percent coverage and $\Delta V$ of less than 0.4 kms$^{-1}$ per satellite per year with 6-12 satellites. On the other hand, the configurations that at the right bottom tend to produce PDOP less than 4 with 95 percent availability. Most of these solutions are composed of a higher number of satellites. The plot of HDOP vs HDOP availability in Figure 11 also follows a similar trend, but HDOP of all solutions range between 1 and 5. The lower HDOP along with the higher coverage availability is achieved with higher number of satellites. The algorithm has also identified some exceptional solutions that can achieve high navigation accuracy with as low as 6 satellites. This outcome points out an interesting fact that better positioning does not depend only on the constellation size, but a superior geometric configuration can make a difference.

The impact of orbital geometry on $\Delta V$ requirements, GPS–LNSS visibility, and PDOP performance has been systematically analyzed to identify configurations that offer optimal trade-offs tailored to diverse mission priorities, ranging from performance and coverage to cost and integrity. Figure 12 illustrates how PDOP availability correlates with orbital elements such as SMA, AOP, and inclination. The analysis reveals that configurations featuring high SMA and inclination, with AOP values within the 80°–140° range, consistently achieve PDOP availability close to 90-100%. This is attributed to the orbital geometry directing the apolune over the lunar south pole, thereby enhancing satellite dwell time over the region. A few solutions exhibiting AOP of greater 140° also possess higher PDOP availability. It is mainly achieved by inserting higher satellites per orbit which compromise the cost factor.

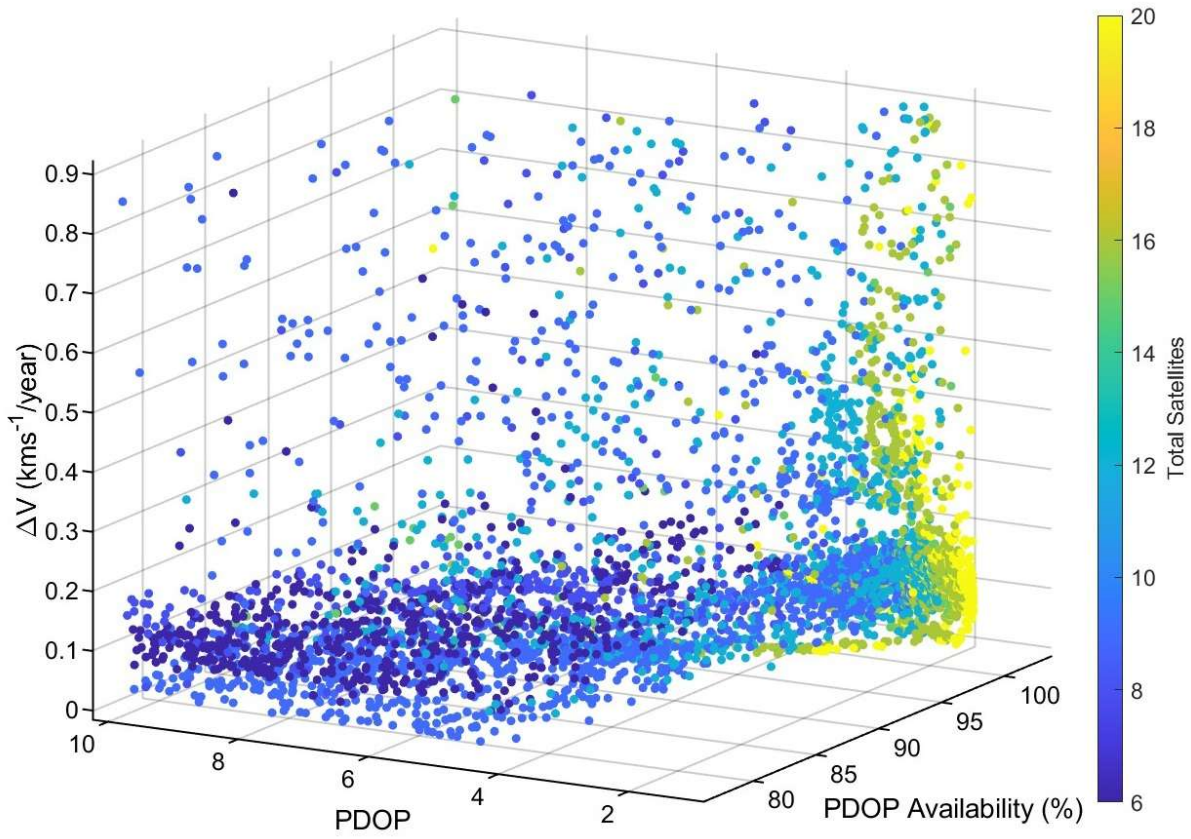

**Figure 10:** LNSS solutions in four-objective space corresponding to PDOP, PDOP availability and station-keeping $\Delta V$ in km$^{-1}$ per satellite per year (Total satellites as color-coded).

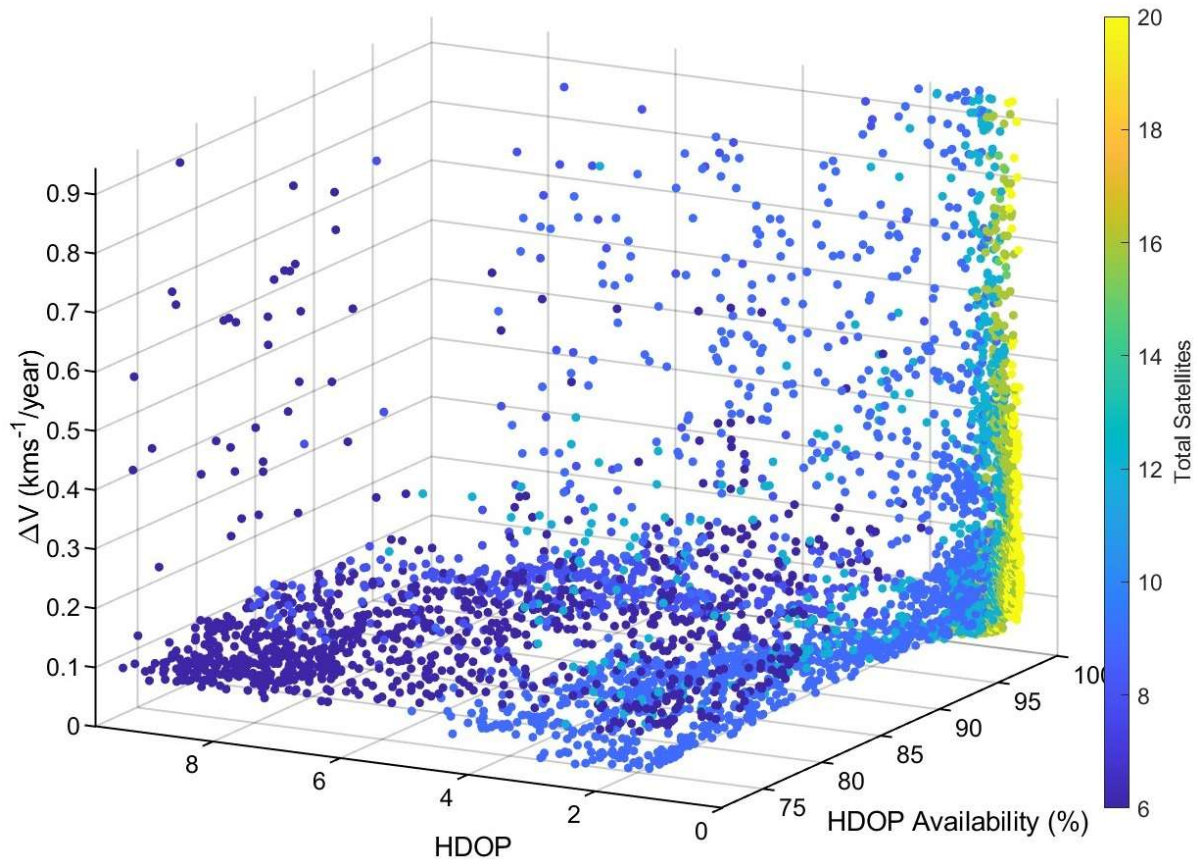

**Figure 11:** LNSS solutions in four-objective space corresponding to HDOP, HDOP availability and station-keeping $\Delta V$ in km$^{-1}$ per satellite per year (Total satellites as color-coded).

In contrast, Figure 13 presents the $\Delta V$ levels of various solutions as a function of SMA, AOP, and eccentricity. It is evident that low SMA orbits generally incur higher $\Delta V$, while highly eccentric orbits are also maneuver-intensive. Notably, when elliptical orbits are geometrically optimized, they can provide robust regional coverage (particularly over the lunar south pole) using a relatively small constellation. The most efficient $\Delta V$ performance is observed in a large cluster of orbits characterized by high eccentricity, large SMA, and AOP values greater than 80°-160°. Additionally, few solutions in the lower right region, featuring low eccentricity and high SMA, also shows favorable $\Delta V$ performance. This is because at higher altitudes, the influence of irregular lunar gravity harmonics diminishes, even in non-frozen orbits, making third-body perturbations the dominant factor, which are comparatively easier to manage. However, the $\Delta V$ efficiency observed in both clusters may come at the cost of requiring a larger constellation size, as reflected in the PDOP availability trends shown in Figure 12.

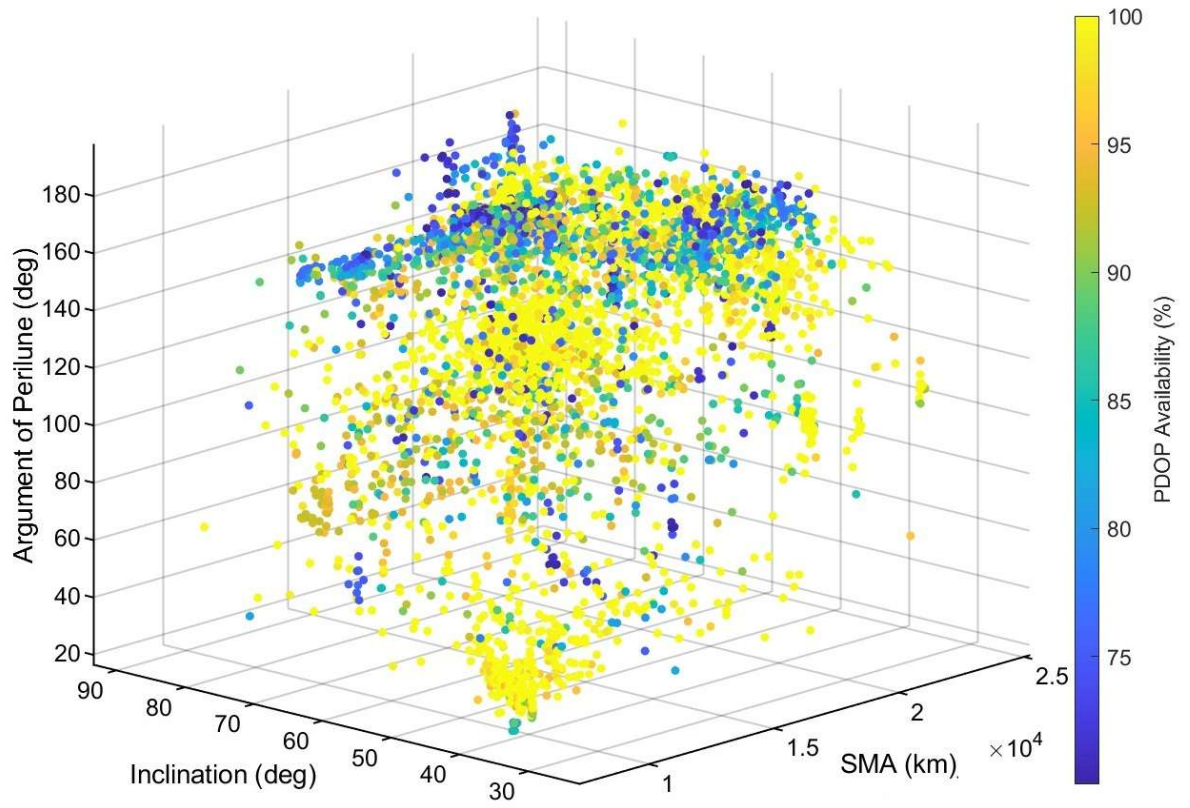

**Figure 12:** Illustration of the relationship between SMA, AOP and inclination with PDOP availability for LNSS solutions in 4-objective space.

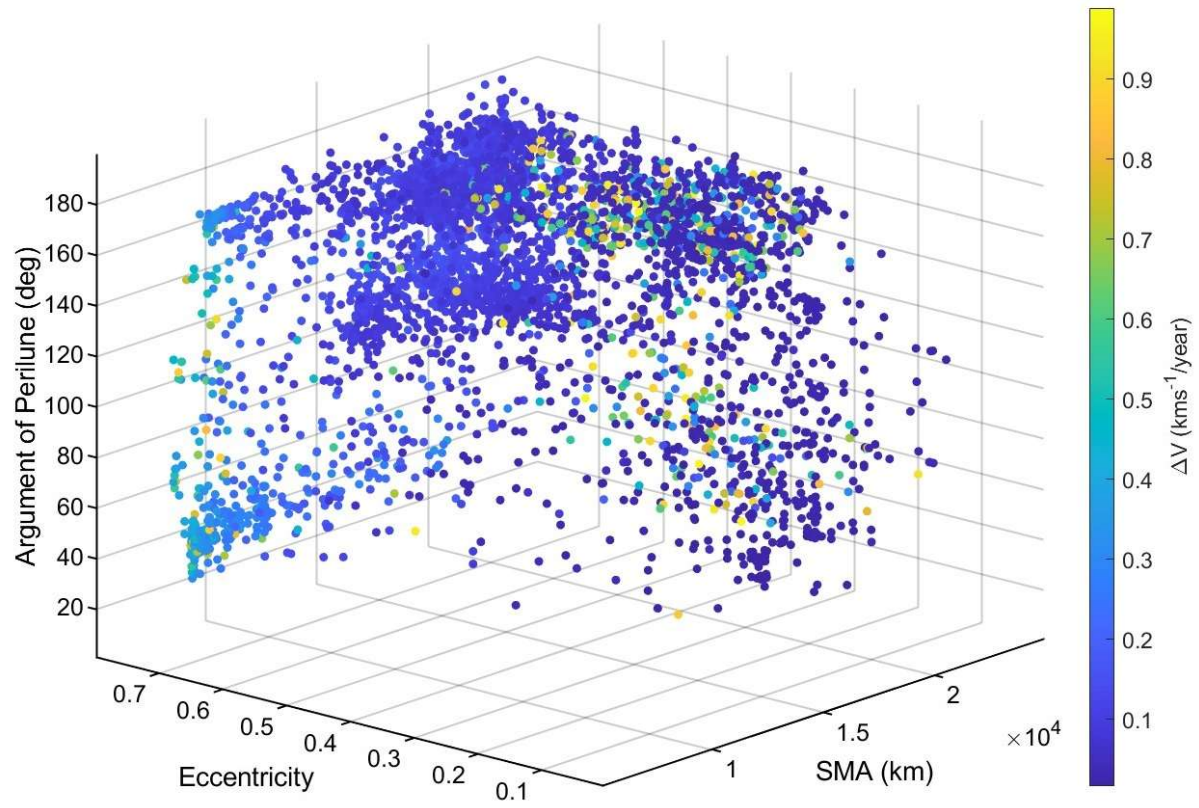

**Figure 13:** Illustration of the relationship between SMA, AOP and Eccentricity with station-keeping $\Delta V$ (km$^{-1}$ per satellite per year) for LNSS solutions in 4-objective space.

The orbital orientation as defined by inclination and AOP has a profound effect on GPS–LNSS visibility and PDOP, driven by geometric alignment. As shown in Figure 14, the solutions at the bottom right occupying small SMA, inclination and AOP yield the poorest PDOP performance. Conversely, configurations at top left occupying 80°-160° AOP, inclinations of 40°-60° and SMA above 20,000 km offer superior and PDOP characteristics. This multi-dimensional analysis highlights a fundamental trade-off in orbit design: the configurations that optimize navigation performance often conflict with those that minimize $\Delta V$ or enhance GPS integration. The optimization framework, therefore, seeks solutions that strike a careful balance among competing objectives, providing mission designers with a spectrum of viable options aligned with specific mission objective priority, e.g. higher navigation accuracy vs GPS accessibility, etc. The key orbital elements of the LNSS solutions is depicted in Figure 15 which shows that most of the solutions comprise highly elliptical orbits (0.6 > e > 0.7) with various combinations of SMA, AOP and inclination that ensure higher satellite visibility over the south pole. This shows that the optimizer has identified the efficient orbital configurations that fulfill the mission objectives.

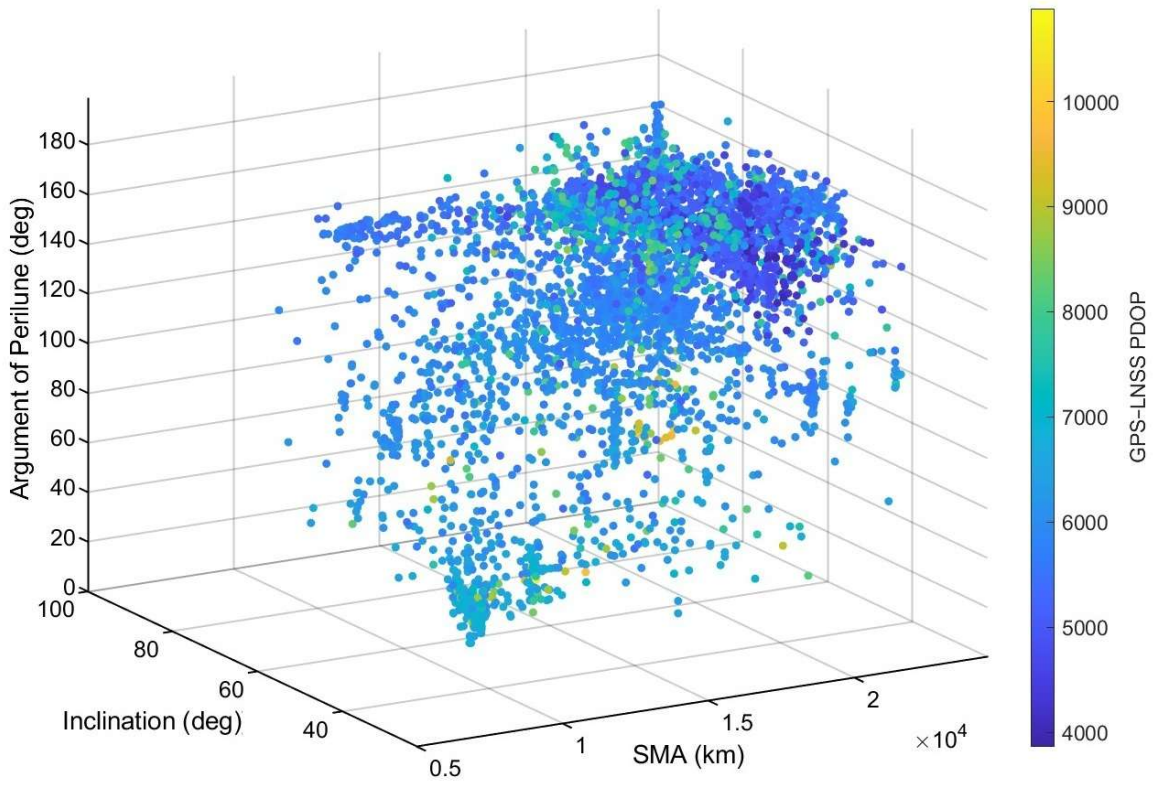

**Figure 14:** Illustration of the relationship between SMA, AOP and inclination with GNSS-LNSS PDOP for LNSS solutions in 4-objective space.

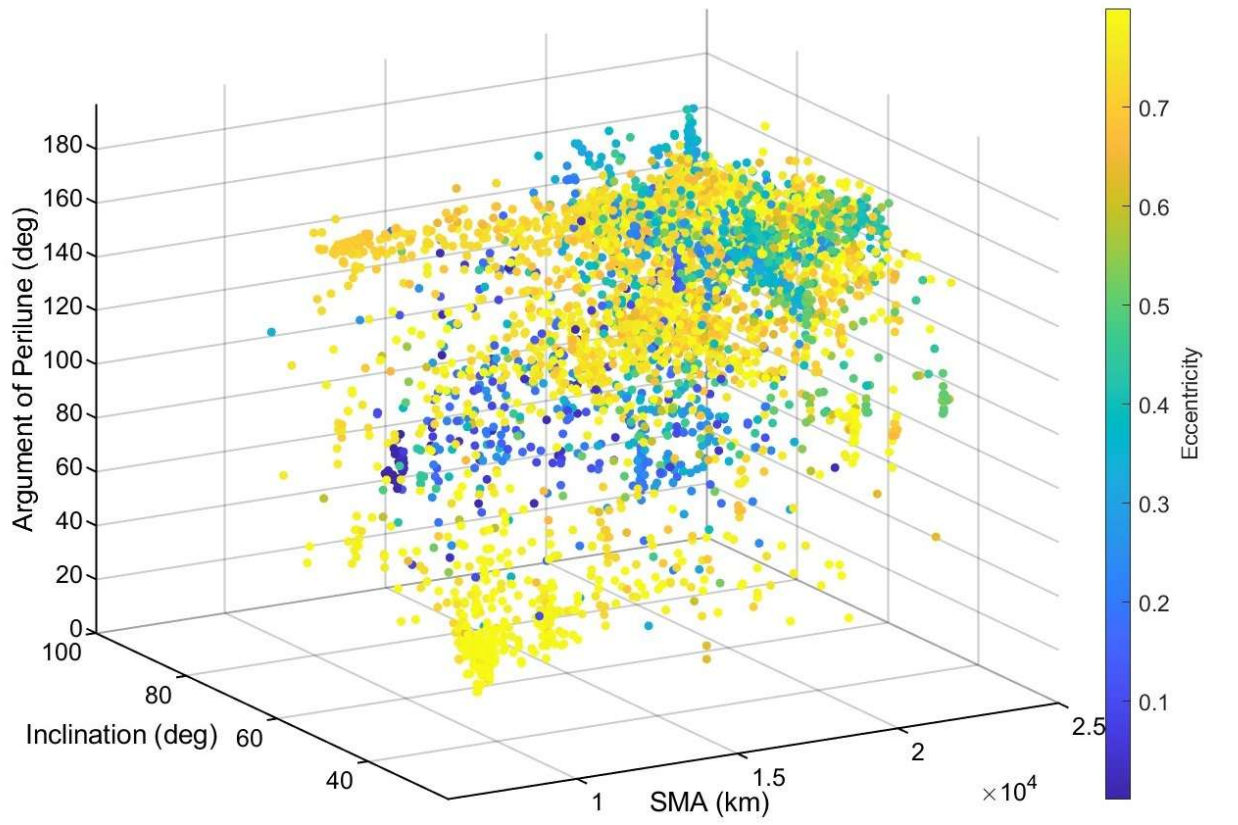

**Figure 15:** Orbital elements representation of the LNSS solutions in 4-objective space.

**Table 6:** The design and objective values of 14 selective LNSS configurations searched by the MOGA algorithm with PDOP and HDOP availability ≥ 90%.

| *ID* | *a* *km* | *e* | *ω* *deg* | *i* *deg* | $N_p$ | $N_S$ | $N_{sat}$ | $\overline{PDOP_{3\sigma}}$ | $PDOP_{avail}$ *%* | $\overline{HDOP_{3\sigma}}$ | $HDOP_{avail}$ *%* | $\Delta V_{Total}$ *kms⁻¹ per satellite per year* | $\Delta V_{Opt}$ *kms⁻¹ per satellite per year* | $PDOP_{GPS-LNSS}$ |
|---|---|---|---|---|---|---|---|---|---|---|---|---|---|---|
| 1 | 20243 | 0.69 | 114 | 754 | 2 | 3 | 6 | 4.3 | 100 | 2.8 | 100 | 0.11 | 0.10 | 5860 |
| 2 | 19308 | 0.77 | 138 | 68.6 | 2 | 3 | 6 | 4.4 | 94 | 3.3 | 94 | 0.17 | 0.16 | 4917 |
| 3 | 15420 | 0.74 | 137 | 79.9 | 2 | 4 | 8 | 4.9 | 100 | 4.5 | 100 | 0.37 | 0.35 | 5506 |
| 4 | 17761 | 0.71 | 98 | 82 | 2 | 4 | 8 | 4.9 | 100 | 3.2 | 100 | 0.10 | 0.09 | 5889 |
| 5 | 19928 | 0.55 | 140 | 38 | 3 | 3 | 9 | 3.5 | 94 | 1.2 | 95 | 0.07 | 0.06 | 3973 |
| **6** | **21225** | **0.67** | **106** | **81** | **3** | **3** | **9** | **3.4** | **100** | **1.5** | **100** | **0.10** | **0.09** | **5782** |
| 7 | 19905 | 0.63 | 142 | 45.9 | 3 | 4 | 12 | 2.3 | 100 | 0.9 | 100 | 0.09 | 0.08 | 3862 |
| 8 | 7447 | 0.72 | 109 | 56.4 | 4 | 3 | 12 | 3.0 | 100 | 0.9 | 100 | 0.55 | 0.44 | 5940 |
| 9 | 12687 | 0.78 | 124 | 41.5 | 5 | 3 | 15 | 2.7 | 100 | 0.7 | 100 | 0.21 | 0.18 | 5033 |
| 10 | 17686 | 0.78 | 114 | 71.9 | 5 | 3 | 15 | 2.2 | 100 | 0.9 | 100 | 0.17 | 0.16 | 5796 |
| 11 | 10539 | 0.78 | 90 | 50.2 | 4 | 4 | 16 | 4.1 | 100 | 0.7 | 100 | 0.84 | 0.73 | 5878 |
| 12 | 23502 | 0.71 | 120 | 47.8 | 4 | 4 | 16 | 2.0 | 100 | 0.7 | 100 | 0.14 | 0.12 | 5279 |
| 13 | 9433 | 0.78 | 83 | 52.8 | 5 | 4 | 20 | 3.1 | 100 | 0.6 | 100 | 0.35 | 0.29 | 6218 |
| 14 | 22989 | 0.71 | 84 | 73.7 | 5 | 4 | 20 | 2.6 | 100 | 0.9 | 100 | 0.09 | 0.07 | 6013 |

A few selective Pareto-front solutions are depicted in Table 6, which shows the orbital geometry (decision variables) and the objective functions within constraints of PDOP and HDOP availability of 90% and a station-keeping $\Delta V \leq 1$ $kms^{-1}$ per satellite per year. Although all the selected solutions define balanced trade-offs, however, solution 6 is the best in terms of positioning capability with less than ten satellites. Likewise, solutions 1 and 2 constitute the least number of satellites and 7 provides the best LNSS-GPS geometry for that lowers PDOP. It must be noted that there are several configurations that fulfill the mission design objectives. These solutions are just for the reference to demonstrate the performance metrics, the constellation size, orbital maintenance and GPS integration for different orbital configurations.

A direct comparison of these results cannot be made with the LNSS optimization as proposed in literature because those studies are based on global lunar coverage. Similarly, the $\Delta V$ computation is based on a different yet high accuracy model and GPS-LNSS PDOP is not included in those studies. However, the results of this study visibly imply that quality positioning in the targeted region can be achieved with significantly small LNSS constellation (6-9 satellites) with an optimized orbital configuration. The main reason for the optimizer to find these cost-efficient solutions is that minimization of constellation size is one of the primary objective functions. The configuration of ID 5 LNSS solution is presented in Figure 16 as a sample for the purpose of visualization.

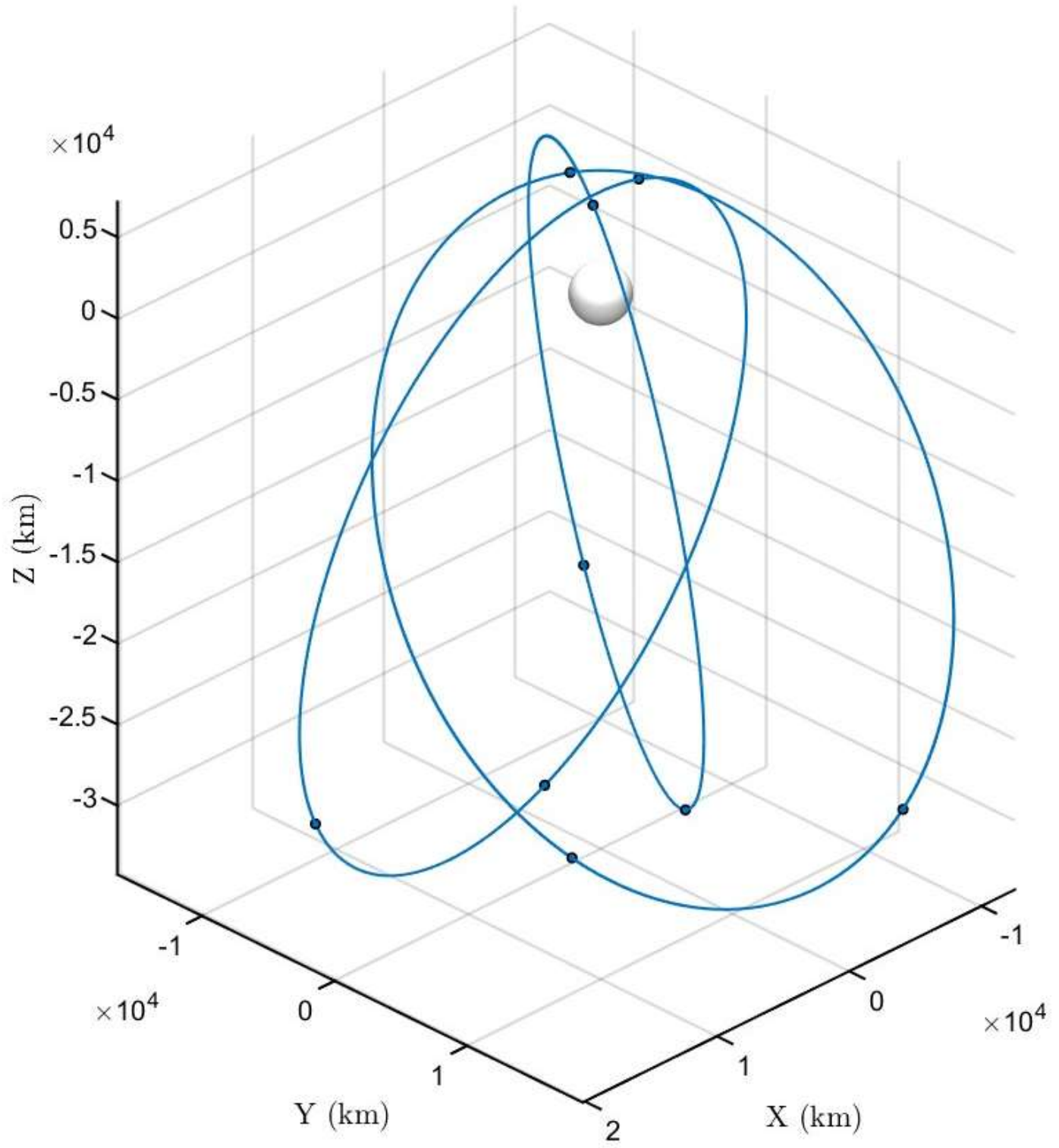

**Figure 16:** Orbital configuration visualization of ID 5 LNSS solution in MCI frame.

# 7 CONCLUSION

In this article, we have developed and validated a comprehensive multi-objective optimization framework for the design of a dedicated Lunar Navigation Satellite System with more comprehensive mission objectives as compared to literature. This approach tightly integrates a high-fidelity numerical propagator containing realistic perturbations with a semi-analytical continuous-thrust based station-keeping $\Delta V$ model. This model drives the orbit evolution and then thrust accelerations for orbital drift correction using GVEs, which is a more precise approach compared to the prior studies. The thrust profiles are then further refined via a small-scale NLP that adaptively optimizes the magnitude and pointing angles of the thrust vector, to explore practical station-keeping budget management in a propellent constraint scenario.

The LNSS navigation performance metrics have been modelled through PDOP and HDOP over a south pole observer grid and assess Earth–GPS integration through detailed line-of-sight, occultation and RF link-budget analyses. A standalone ELFO-based constellation is propagated to better demonstrate computational process of objective function variables. Subsequently, full-scale optimization is executed with NSGA-II within MATLAB environment to explore a six-dimensional decision space (orbital parameters). The resulting Pareto fronts reveal the trade-offs among constellation size, navigation accuracy, GPS-LNSS interoperability and $\Delta V$ budget. This full integrated optimization simulation run identified numerous compact constellations achieving PDOP < 10 with ≥ 70 % availability, GPS-LNSS PDOP $\lesssim$ 5000 and $\Delta V \lesssim 1$ kms$^{-1}$ per satellite per year using numerous combinations of orbital configurations. The results suggest that there exist numerous non-frozen elliptical obits that can achieve acceptable navigation accuracy with low number of satellites and adequate $\Delta V$ needs. The relationship of orbital arrangement with the objective functions also suggests that the orbital geometry best suited for the navigation performance with small constellation may be expansive in terms of Earths GPS accessibility and/or the $\Delta V$ budget and vice versa. Therefore, the optimizer has defined balanced trade-offs as reflected in the best configurations presented in Table 6.

This framework accounts for more multi-dimensional objectives and a rigorous semi-analytical $\Delta V$ computation model as compared to literature. Moreover, even if this paper focused only on the south pole, the design space has identified solutions with lower number of the satellites as compared to reference [9] and [10]. As no prior research has explored the GPS-LNSS integration, and the methodology for station-keeping $\Delta V$ is also not the same thus comparison cannot be made. The proposed framework strikes a balance between computational tractability and physical fidelity, yielding actionable design candidates for future polar LNSS deployments.

Going forward, incorporating on-board orbit determination and time-transfer algorithms, robustness to launch and injection errors, heterogeneous propulsion architectures, and autonomous maneuver planning / implementation will further enrich the design space and support robust and cost-effective lunar navigation infrastructures. The GPS-LNSS PDOP analysis suggests that relying on one GNSS system (e.g. GPS only) is insufficient for effective integration. The future studies on LNSS design and operational aspects should integrate other GNSS systems for further enhancements.

# 8 Credit Authorship Contribution Statement

**Rameez A. Malik:** Conceptualization of this study, Methodology, Programming, Writing - Original draft preparation; **Yang Yang:** Conceptualization, Supervision and Editing.

# 9 Declaration of Generative AI in Scientific Writing

*During the preparation of this work the author(s) used Chat GPT in order to seek assistance in writing process. After using this tool/service, the author(s) reviewed and edited the content as needed and take(s) full responsibility for the content of the publication.*

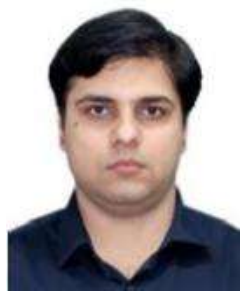

**RAMEEZ A. MALIK** received his B.S. and M.Phil degree in space sciences from University of the Punjab, Lahore, Pakistan, in 2015 and 2018, respectively. Currently, he is a PhD scholar at University of New South Wales, Sydney, Australia. His research focuses on the design of lunar navigation system with emphasis on dedicated lunar constellation optimization, lunar orbit estimation and time transfer from the Earth positioning systems to future lunar navigation system. He has over five years of experience in space flight dynamics and space operations. He is also a reviewer of IET Radar, Sonar and Navigation, ION Navigation, and Mathematical Problems in Engineering.

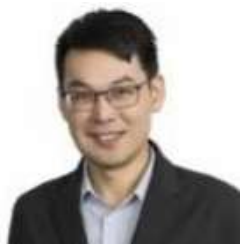

**YANG YANG** received the Ph.D. degree from Northwestern Polytechnical University, Xi'an, China. He is currently a Space Engineering Lecturer with the School of Mechanical and Manufacturing Engineering, The University of New South Wales (UNSW), Sydney, NSW, Australia. His research interests include advanced estimation, spacecraft navigation, space situational awareness, and GNSS positioning and navigation.